\documentclass[preprint,review,12pt]{elsarticle}
\usepackage{subfig}
\usepackage{caption}
\usepackage{epsfig,multirow,multicol}
\usepackage[colorlinks]{hyperref}
\usepackage{amssymb}
\usepackage{amsthm}
\usepackage{amsmath}
\usepackage{graphicx,hyperref}
\usepackage{epstopdf}
\usepackage{bm}             
\usepackage{lineno}
\allowdisplaybreaks
 \newcommand{\be}{\begin{equation}}
 \newcommand{\ee}{\end{equation}}
 \newcommand{\bea}{\begin{eqnarray}}
 \newcommand{\eea}{\end{eqnarray}}
 \newcommand{\bwt}{\begin{widetext}}
 \newcommand{\ewt}{\end{widetext}}

 \allowdisplaybreaks

\journal{Journal of Magnetism and Magnetic Materials}

\begin{document}

\begin{frontmatter}

\title{Temperature and Field Dependence of Ferromagnetic Magnon in Monolayer Honeycomb Spin Lattice}

\author{Niem T.~Nguyen$^{1}$, Giang H.~Bach$^{1}$, Thao H.~Pham$^{2}$, Huy D.~Nguyen$^1$, Oanh T.~K.~Nguyen$^{3}$, Cong T.~Bach$^{1}$} 

\address{$^{1}$Faculty of Physics, University of Science, Vietnam National University,\\
    334 Nguyen Trai street, Thanh Xuan, Hanoi, Vietnam\\
	$^2$Hue University, 34 Le Loi, Hue, Vietnam \\
	$^3$Electric Power University,
	235 Hoang Quoc Viet street, Bac Tu Liem, Hanoi, Vietnam \\
	}

\begin{abstract}
Temperature and field dependence of collective spin excitations or magnon in monolayer honeycomb spin lattices is investigated using an anisotropic exchange XZ-Heisenberg model in an external field. Magnetic phase transition in the presence of the transverse field is the spin reorientation (SR) transition with magnon intensity existing above the SR temperature. The transverse field either decreases or sustains the spin-wave intensity in the temperature region below or above the SR temperature, respectively. The gap of the zero-momentum low-energy magnon branch closes at the SR transverse field, which is the critical quantum phase transition field at zero temperature. The application of the model to a two-dimensional CrI$_3$ explains the existence of the zero-momentum magnon mode above the Curie temperature and shows the suitable values of the exchange parameters compared with the DFT calculations. The estimated magnon velocity near the Dirac point in this material is about 1.74 km/s. 

\end{abstract}


\end{frontmatter}

\section{Introduction}
Dirac materials having elementary excitation spectra similar to the Dirac electronic quasiparticle in graphene are intensively investigated for the last decade \cite{Balatsky2014}. Among the Dirac fermion, Dirac bosons such as magnon in honeycomb structural magnets have drawn great attention to both fundamental and practical research. It is shown that due to the spatial symmetry of the honeycomb spin-lattice, the magnon energy spectra have a Dirac-like dispersion either around the K and K’ points of the Brillouin zone (BZ) for the ferromagnetic (FM) magnon or around the center of the BZ for the antiferromagnetic (AF) magnon  \cite{Balatsky2016,Balatsky2018}. It is also proposed that competing interactions in the FM structure induce Dirac and Weyl points while Dirac nodes relate to the magnetic structure but not to the overall crystal symmetry \cite{Boyko2018}. 

Based on graphene-like two-dimensional (2D) materials, Van der Waals crystal chromium triiodide CrI$_3$, one of the typical 2D magnets, exhibits ferromagnetic order when the size of the crystal is down to a single layer \cite{Huang2017, Novo2019}. Acoustical and optical magnon branches are recognized in the mono-layer CrI$_3$ using magneto-Raman spectroscopy \cite{Cenker2021}. Temperature dependence of the two branches of the zero-momentum magnon (k=0) has been experimentally investigated in the 2D honeycomb ferromagnet CrI$_3$ \cite{He2018} and analyzed by the anisotropic Heisenberg model. It is indicated that the magnon energy reduces with increasing temperature below the phase transition temperature (T$\rm_C$), whereas it remains finite, and nearly temperature-independent above this. In the framework of the XZ-Heisenberg model with the transverse field (TF), the spin wave in the monolayer square spin-lattice can exist above T$\rm _C$ and displays a weak temperature dependence \cite{Nguyen2018}. The phase transition temperature in the presence of the TF can be interpreted as the spin reorientation (SR) temperature following Ref.~\cite{Nolting2005}. This transition due to the TF at zero temperature belongs to quantum phase transition (QPT) where the TR plays a role of tuning parameters which probably are pressure, doping fraction, in-plane stress, etc. \cite{Scott2015}.

The longitudinal magnetic-field-induced quantum phase transition was experimentally explored in the CrI$_3$ bulk with honeycomb spin structure, where antiferromagnetic surface layers transformed to a ferromagnetic (FM) state at the critical field of about 2 T \cite{Li2020}. The magnetic phase transitions in the single CrI$_3$ layer induced by stress due to in-plane lattice deformation are theoretically examined in Ref.~\cite{Yaz2020}. The temperature-independent magnon intensity above the phase transition temperature \cite{He2018} was particularly described by the spin-wave theory in the XZ-Heisenberg model with the TF which mimics the in-plane stress \cite{Nguyen2018}. According to our knowledge, the finite temperature and transverse field behaviors of the honeycomb spin-lattice FM magnon have received relatively modest attention.

For this reason, we aim to deeply understand the temperature and the TF behavior of the monolayer honeycomb spin-lattice magnons, including the QPT case. Our results bring specific features of magnon energy in the CrI$_3$ and other 2D systems observed in experiments. The paper is organized as follows: the next part describes the model and the calculation method. In Section III, we analyze the FM magnon spectra in the monolayer honeycomb spin-lattice and compare our model parameters with experimental calculations. We will summarize our results in the conclusion part. 

 \section{XZ-Heisenberg model and Green function}
 
 \subsection{Hamiltonian of the XZ- Heisenberg model for monolayer films}
 \begin{figure}
\centering
    \subfloat[\label{fig:1a}]{\includegraphics[height=2.in,width=2.5in]{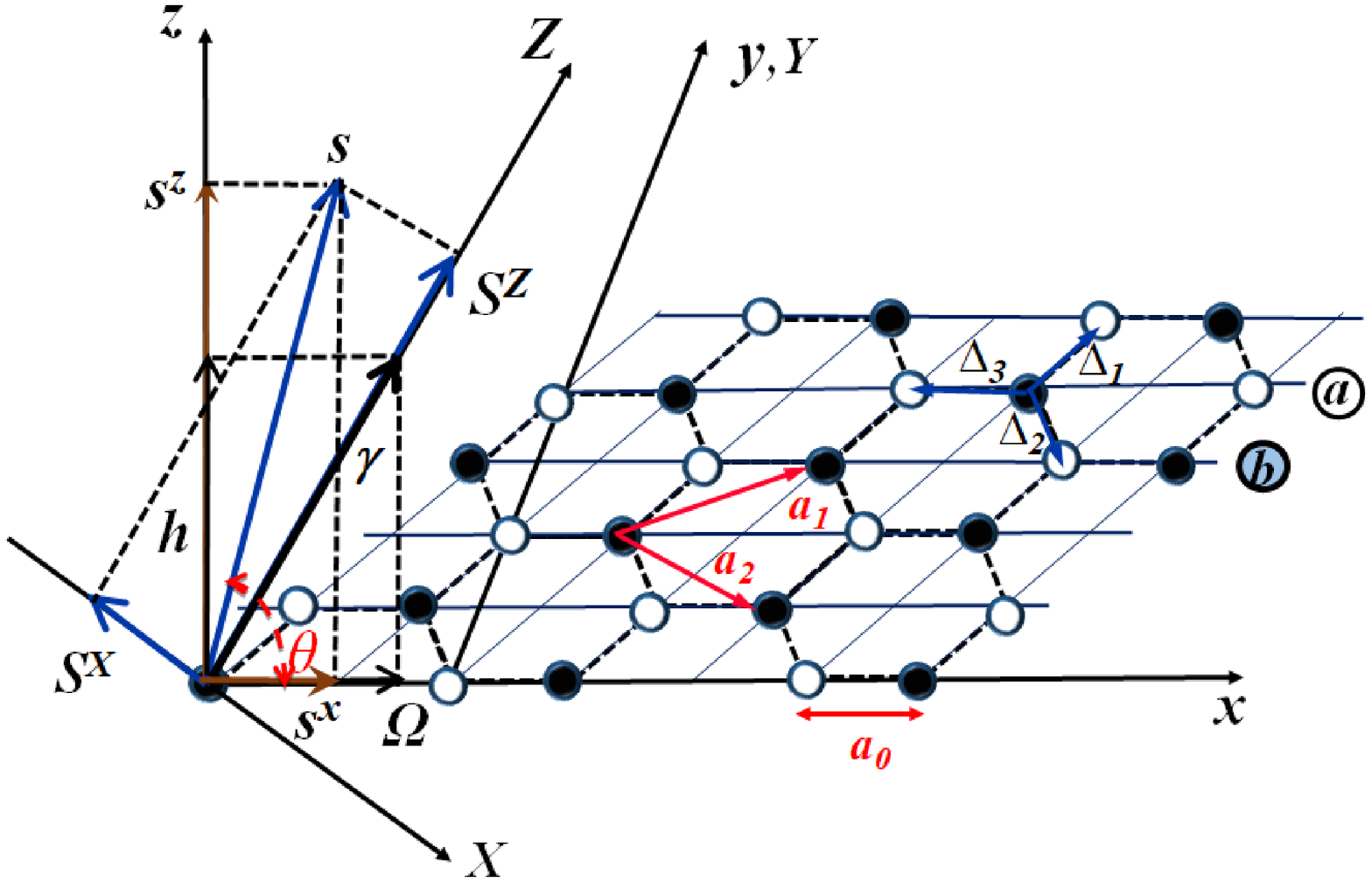}}%
    \hspace{1cm}
    \subfloat[\label{fig:1b}]{\includegraphics[height=2.in,width=2.in]{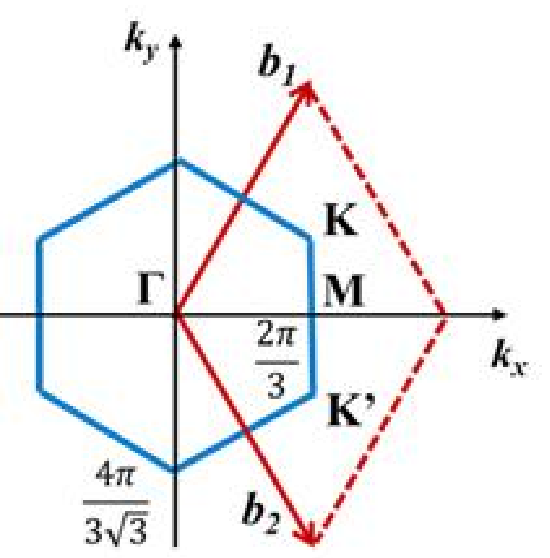} }
 \caption{Orientation of a spin in the honeycomb spin-lattice relates to the crystallographic (xyz) and rotated (XYZ) coordinate systems \protect\subref{fig:1a} and the first Brillouin zone \protect\subref{fig:1b}.}
 \label{Fig1}%
\end{figure}
 
We consider a monolayer honeycomb spin-lattice consisting of 2N spins residing in two triangular $a$ and $b$ spin sub-lattices shown in Fig.~\ref{Fig1}\protect\subref{fig:1a}. Position vectors of these spins in the sub-lattices are denoted by two-dimensional lattice vectors $\bf j$ and $\bf g$. Three nearest-neighbors (NN) of an $a-$spin at site j are $b-$spins defined by three vectors $\bm \Delta$ which are ${\bm \Delta_1}={a_0 \over 2}(3,\sqrt3)$, ${\bm \Delta_2}={a_0 \over 2}(3,-\sqrt3)$, ${\bm \Delta_3}={a_0 \over 2}(-1,0)$ with the length of the hexagonal edge $a_0$. Similarly, three NN of a $b-$spin at site g are $a-$spins positioned by vectors $-\bm \Delta$. The six next-nearest-neighbors (NNN) of the $a$($b$)-sub-lattice are the same spin types defined by vectors $ {\bm \rho}_1={a_0 \over 2}(3,\sqrt3) =-{\bm \rho}_4$, ${\bm \rho}_2={a_0 \over 2}(3,-\sqrt3)=-{\bm \rho}_5$, ${\bm \rho}_3={a_0 \over 2}(0,-\sqrt{3})=-{\bm \rho}_6$. 

The Hamiltonian of the  XZ$-$Heisenberg model for the honeycomb monolayer spin-lattice in the external fields is written by
\bea
 	H = -  \sum_{\bf j}(h_0s^z_{a {\bf j}} + \Omega_0s^{x}_{a\bf j}) -\sum_{\bf g} (h_0s^z_{b\bf g} +\Omega_0s^x_{b\bf g})
- {1 \over 2} \sum_{{\bf j},{\bm \Delta}} (Js^{z}_{a \bf j}s^{z}_{b, {\bf j}+ {\bm\Delta}} + Ls^{x}_{a\bf j}s^{x}_{b, {\bf j}+{\bm\Delta}})  \nonumber\\ 
	-  {1 \over 2} \sum_{{\bf g},{\bm \Delta}} (Js^{z}_{b \bf g}s^{z}_{a, {\bf g}-{\bm\Delta}} + Ls^{x}_{b\bf g}s^{x}_{a, {\bf g}-{\bm\Delta}})  
	-  {1\over 2} \sum_{\bf j,{\bm \rho}} J^{'}(s^{z}_{a\bf j}s^{z}_{b,{\bf j}+ {\bm \rho}} + s^{x}_{a\bf j}s^{x}_{b,{\bf j}+{\bm \rho}}) \nonumber \\ 
	-  {1 \over 2} \sum_{\bf g,\bm \rho} J^{'} (s^{z}_{b\bf  g}s^{z}_{a, \bf g-\bm \rho} + s^{x}_{b\bf g}s^{x}_{a,\bf g- \bm\rho} )
	\label{Eq1}
\eea
Here $ h_0$, $\Omega_0$ are the external longitudinal and transverse field intensity given in the energy unit. The components of sub-lattice spin operators in the crystallographic $xyz$ frame, where the $z–$axis is perpendicular to the honeycomb spin-lattice plane, are denoted by $s^{z}_{a\bf j}$, $ s^{x}_{a\bf j}$, $ s^{z}_{b\bf g}$, $ s^{x}_{b\bf g}$. $J$ and $L$ are NN exchange parameters between spin-pair components along the out-of-plane $z$ and in-plane $x$ directions, respectively. Since long-range order generally does not exist in a 2D isotropic exchange Heisenberg model \cite{Mermin1966}, the exchange parameters are chosen as $J \neq L$. $J’$ is the NNN exchange between spins of the same types.

Using the mean-field approximation (MFA), we rewrite the Hamiltonian (\ref{Eq1}) in terms of spin fluctuation operators $ \delta s^{z}_{a(b)\bf j} = s^{z}_{a(b)\bf j} - m_{az(bz)} $, 
\be
H =   H_0 + H_1  \,, 
\ee
where 
\bea
H_0 =&  3N[ Jm_{az}m_{bz} + Lm_{ax}m_{bx} + J'(m^2_a +m^2_b)] \nonumber \\
- &  \sum_{\bf j} (h_a s^z_{a\bf j} + \Omega_a s^{x}_{a\bf j}) - \sum_{\bf g} (h_b s^z_{b\bf g} + \Omega_b s^{x}_{b\bf g}) \,, 
\label{Eq2a}
\eea
\bea
 H_1 = &  -{ 1\over 2} \sum_{{\bf j}{\bm\Delta}} (J \delta s^{z}_{a\bf j} \delta s^{z}_{b,{\bf j}+{\bm\Delta}} + L\delta s^{x}_{a\bf j} \delta s^{x}_{b,{\bf j}+{\bm\Delta}} ) \nonumber \\
-&  { 1\over 2} \sum_{{\bf g}{\bm \Delta}} (J\delta s^{z}_{a\bf g} \delta s^{z}_{b,{\bf g}-{\bm \Delta}} + L\delta s^{x}_{a\bf g} \delta s^{x}_{b,{\bf g}-{\bm \Delta}} ) \nonumber \\ 
-&  { 1\over 2} \sum_{\bf j \bm\rho} J'(\delta s^{z}_{a\bf j} \delta s^{z}_{b,\bf j+ \bm\rho} + \delta s^{x}_{a\bf j} \delta s^{x}_{b,\bf j+\bm \rho} ) \nonumber \\
-& { 1\over 2} \sum_{\bf g \bm \rho} J'(\delta s^{z}_{a\bf g} \delta s^{z}_{b,\bf g-\bm \rho} + L\delta s^{x}_{a\bf g} \delta s^{x}_{b,\bf g-\bm\rho} ), 
\label{Eq2b}
\eea
and 
\bea
 h_{a(b)} = &  h_0 + 3Jm_{bz(az)} + 6J'm_{az(bz)}, \nonumber \\
\Omega_{a(b)} =&  \Omega_0 + 3Lm_{bx(ax)} + 6J'm_{ax(bx)} \,.
\label{Eq3}
\eea 
Here $ m_{\nu x}=<s^x_\nu>$, $ m_{\nu z}=<s^z_\nu>$ with $\nu= a,b$ are thermodynamic average of the spin-moment components per site and $m_\nu =\sqrt{m^2_{\nu x} + m^2_{\nu z}}$, and $ <...> = {\rm Tr}(e^{-\beta H}... )/{\rm Tr}(e^{-\beta H})$. The formula (\ref{Eq3}) produces the longitudinal and transverse components, $h_\nu$ and $\Omega_\nu$, of the net field acting on the $\nu$-sub-lattice. The Hamiltonian $H_0$ is diagonalized using the transformation, 
\bea
 s^x_{\nu \bf j} =& {h_{\nu} \over \gamma_\nu} S^X_{\nu \bf j}  + {\Omega_{\nu} \over \gamma_\nu} S^Z_{\nu \bf j}\,, \nonumber\\
 s^z_{\nu \bf j} =& -{\Omega_{\nu} \over \gamma_\nu} S^X_{\nu \bf j}  + {h_{\nu} \over \gamma_\nu} S^Z_{\nu \bf j} \,.
\label{Eq4}
\eea
 During the transformation, spin fluctuation operators $\delta s^\alpha_{\nu \bf j}$ ($\rm \alpha=x,z$) are transformed to new operators
$ \delta S^\alpha_{\nu \bf j} = S^\alpha_{\nu \bf j} - <S^\alpha_{\nu}>$, ($\alpha=X,Z$). Performing Fourier transformation $\delta S^\alpha_{\nu \bf j}=N^{-1/2}\sum_{\bf k} \delta S^\alpha_{\nu}({\bf k}) e^{i{\bf kj}}$ with two-dimensional wave vector $\bf k$ defined in the first Brillouin zone (shown in Fig.~1(b)), we rewrite the Hamiltonian (\ref{Eq2a}) and (\ref{Eq2b}) as 
\bea
H_0 = 3N \{ Jm_{az}m_{bz} + Lm_{ax}m_{bx} + J'(m^2_a + m^2_b ) \} -\sum_{\nu j} \gamma_\nu S^z_{\nu j} ,\qquad
\label{Eq7}
\eea
and
\bea
H_1 = -{1\over 2} \sum_{\bf k} \sum_{\lambda \lambda'} I_{\lambda \lambda'} ({\bf k}) \delta S^{\lambda}({\bf k})
\delta S^{\lambda'}({\bf -k}),
\label{Eq8}
\eea
where $\lambda$ and $\lambda'$ run from 1 to 4 meaning $1\equiv aX$, $2\equiv aZ$, $3\equiv bX$, $4\equiv bZ$. The anisotropic exchange interaction is a Hermitian matrix defined by 
\begin{equation}
	\hat{I}({\bf k}) =
	\begin{pmatrix}
		\xi_{2\bf k} & 0&  C\xi^{\ast}_{1\bf k} & D \xi^{\ast}_{1\bf k} \\
		0& \xi_{2\bf k} & \tilde{D}\xi^{\ast}_{1\bf k} &  \tilde{C}\xi^{\ast}_{1\bf k} \\ 
	C\xi_{1\bf k} &  \tilde{D}\xi_{1\bf k} & \xi_{2\bf k} &0 \\
	D\xi_{1\bf k} & \tilde{C}\xi_{1\bf k} & 0&\xi_{2\bf k}	
\end{pmatrix}\,,
\label{Eq9}
\end{equation}
where 
\bea
 C= \frac{(J\Omega_a \Omega_b +Lh_a h_b)}{ \gamma_a\gamma_b},\,
 \tilde{C}= {(L\Omega_a \Omega_b +Jh_a h_b) \over \gamma_a\gamma_b},\qquad\\
 D = {(-J\Omega_a h_b +L\Omega_b h_a) \over \gamma_a\gamma_b},\,\, 
 \tilde{D}= {(L\Omega_a h_b - J\Omega_b h_a) \over \gamma_a\gamma_b},\,\qquad
\eea
and 
\bea
\xi_{1\bf k} =& \sum_{\bm \Delta}e^{i{\bf k}{\bm\Delta}}=& e^{-i k_x a_0} (1+2e^{i3k_x a_0/2} {\rm cos}(\sqrt{3}k_ya_0/2)) \, , \nonumber \\
\xi_{2\bf k} =& \sum_{\bm \rho} e^{i\bf k \bm \rho} =&  4 {\rm cos}(3k_xa_0/2){\rm cos}(\sqrt{3}k_ya_0/2) + 2{\rm cos}(\sqrt{3}k_ya_0) \,.  
\label{Eq10}
\eea
\subsection{Green's functions}
The temperature and field dependence of the magnon spectra is obtained from the poles of the imaginary time Green’s function (GF) as carried out in Ref.~\cite{Nguyen2018}. The Green’s function basing on the spin fluctuation operators $ \delta s^{\alpha}_{\nu j}$ ($ \alpha = x,z$; $ \nu= a,b$) given in the crystallographic spin-lattice coordinates and in the Heisenberg picture is written as
\bea
G^{\alpha\alpha'}_{\nu\nu} ({\bf j-j'},\tau_1 -\tau_2) = <\hat{T} \delta \tilde{s}^{\alpha}_{\bf j\nu}(\tau_1) \delta \tilde{s}^{\alpha'}_{\bf j'\nu'}(\tau_2)> \,,
\eea
where
\bea
\delta \tilde{s}^{\alpha}_{\bf j\nu}(\tau) = e^{\tau H} \delta s^\alpha_{\bf \nu j} e^{-\tau H}.
\eea
Taking Fourier transformation of the GF, we have
\bea
G^{\alpha\alpha'}_{\nu\nu'}({\bf j-j'},\tau) = {1\over N} \sum_{{\bf k}\omega} G^{\alpha\alpha'}_{\nu\nu'}({\bf k},\omega) e^{-i{\bf k}({\bf j-j'})-i\omega\tau}\,,
\label{Eq13}
\eea
where $\omega = 2\pi n/\beta$; $n = 0,\pm 1,\pm 2,...$. Denoting $(\bf k,\omega)$ as the three-component wave vector $\bf q$ and putting $\bf R=j-j'$, the Fourier image of the GF writes
\bea
G^{\alpha\alpha'}_{\nu\nu'}({\bf q}) =& <\hat{T} \delta \tilde{s}^{\alpha}_{\bf \nu}({\bf q}) \delta \tilde{s}^{\alpha'}_{\bf \nu'}({\bf q})> \nonumber \\
=& {1\over \beta} \sum_{\bf R} \int_0^{\beta} d\tau G^{\alpha\alpha'}_{\nu\nu'}({\bf R},\tau) e^{i{\bf kR}+i\omega\tau}\,,   
\label{Eq14}
\eea
where
\bea
\delta \tilde{s}^{\alpha}({\bf q}) = {1\over \beta\sqrt{N}} \sum_{\bf j} \int_0^{\beta} \delta \tilde{s}^{\alpha}_{\bf j}({\tau}) e^{i{\bf kj}+i\omega\tau} d\tau\,, \\
\delta \tilde{s}_{\bf j}^{\alpha}(\tau) = {1\over \sqrt{N}} \sum_{\omega,\bf k} \delta \tilde{s}^{\alpha}_{\bf j}({\bf k},\omega) e^{-i{\bf kj}-i\omega\tau} d\tau\,.
\label{Eq16}
\eea
The unitary transformation (\ref{Eq4}) produces the connection between the Fourier images of the GF in the crystallographic and local XZ coordinates, 
\bea
 G^{xx(zz)}_{\nu\nu}({\bf q}) =& {1\over \gamma^2_\nu} \{ \Omega^2_\nu \Gamma^{XX}_{\nu\nu}({\bf q}) + h^2_\nu \Gamma^{ZZ}_{\nu\nu} ({\bf q})  \,\,\,\, \nonumber \\
 \pm & h_\nu \Omega_\nu [\Gamma^{ZX}_{\nu\nu} ({\bf q})+ \Gamma^{XZ}_{\nu\nu} ({\bf q})] \}, \label{Eq21a} \\ 
 G^{xx(zz)}_{ab}({\bf q}) =&  {1\over \gamma_a\gamma_b}\{ \Omega_a \Omega_b \Gamma^{XX}_{ab}({\bf q}) + h_a h_b \Gamma^{ZZ}_{ab} ({\bf q})  \,\,\,\, \nonumber \\
  \pm & [h_a \Omega_b \Gamma^{ZX}_{ab} ({\bf q})+h_b \Omega_a \Gamma^{XZ}_{ab} ({\bf q})] \},\\
 G^{xz}_{ab}({\bf q}) =& {1\over \gamma_a\gamma_b}\{ -h_a \Omega_b \Gamma^{XX}_{ab}({\bf q}) + \Omega_a h_b \Gamma^{ZZ}_{ab} ({\bf q})  \,\,\,\, \nonumber \\
+ & [h_a h_b \Gamma^{XZ}_{ab} ({\bf q})-\Omega_a \Omega_b \Gamma^{ZX}_{ab} ({\bf q})] \}, \\
 G^{zx}_{ab}({\bf q}) =&\rm {1\over \gamma_a\gamma_b}\{ -h_b \Omega_a \Gamma^{XX}_{ab}({\bf q}) + \Omega_b h_a \Gamma^{ZZ}_{ab} ({\bf q})  \,\,\,\, \nonumber \\
+ & [h_a h_b \Gamma^{ZX}_{ab} ({\bf q})-\Omega_a \Omega_b \Gamma^{XZ}_{ab} ({\bf q})] \}.
\label{Eq21}
\eea
The Fourier transform of Green’s function in the local XZ coordinate presented in Eqs.~(\ref{Eq21a})-(\ref{Eq21}) are obtained analogously to Eqs.~(\ref{Eq13})-(\ref{Eq14}) using indexes $\lambda$, $\lambda'$, 
\bea
\rm \Gamma^{\lambda\lambda'} ({\bf j-j'},\tau-\tau')=&\rm <\hat{T} \delta \tilde{S}^{\lambda}_{\bf j}(\tau) \delta \tilde{S}^{\lambda'}_{\bf j'}(\tau')> \,, \\
 \rm \Gamma^{\lambda\lambda'}({\bf q}) =&\rm  <\hat{T} \delta \tilde{S}^{\lambda}({\bf q}) \delta \tilde{S}^{\lambda'}({-\bf q})> \,,\nonumber\\
 = & \rm {1\over \beta} \sum_{\bf R} \int_0^{\beta} d\tau \Gamma({\bf R},\tau) e^{i{\bf kR}+i\omega\tau}.\quad  
 \label{Eq23}
\eea
The GF in Eq.~(\ref{Eq23}) is defined by
\bea
\Gamma^{\lambda\lambda'}({\bf q}) =  { <\hat{T} \delta \tilde{S}^{\lambda}({\bf q}) \delta \tilde{S}^{\lambda'}({-\bf q}) \sigma(\beta)>_0 \over <\sigma(\beta)>_0} \,,
\eea
where $ <...>_0 = {\rm Tr}(e^{-\beta H_0} ...)/{\rm Tr}(e^{-\beta H_0})$ implies the thermodynamic average with the mean-field Hamiltonian in Eq.~(\ref{Eq7}) and $\sigma(\beta)= \hat{T}{\rm exp}[-\int_0^\beta H_1(\tau)d\tau]$ is the scattering matrix. The Green’s function matrix $\hat {\Gamma}({\bf q})$ is presented by, 
\bea
\hat{\Gamma}({\bf q}) =& \begin{pmatrix}
		\hat{\Gamma}_{aa}({\bf q}) &  \hat{\Gamma}_{ab}({\bf q})\\
		\hat{\Gamma}_{ba}({\bf q}) &  \hat{\Gamma}_{bb}({\bf q}) 
	\end{pmatrix} \nonumber \\ =& \begin{pmatrix}
	 \Gamma^{XX}_{aa}({\bf q}) &  \Gamma^{XZ}_{aa}({\bf q}) &  \Gamma^{XX}_{ab}({\bf q}) &  \Gamma^{XZ}_{ab}({\bf q}) \\
	 \Gamma^{ZX}_{aa}({\bf q}) &  \Gamma^{ZZ}_{aa}({\bf q}) &  \Gamma^{ZX}_{ab}({\bf q}) & \Gamma^{ZZ}_{ab}({\bf q}) \\
	 \Gamma^{XX}_{ba}({\bf q}) &  \Gamma^{XZ}_{ba}({\bf q}) & \Gamma^{XX}_{bb}({\bf q}) &  \Gamma^{XZ}_{bb}({\bf q}) \\
	\Gamma^{ZX}_{ba}({\bf q}) &  \Gamma^{ZZ}_{ba}({\bf q}) & \Gamma^{ZX}_{bb}({\bf q}) &  \Gamma^{ZZ}_{bb}({\bf q}) 
\end{pmatrix}.
\eea

In Ref.~\cite{Nguyen2018}, we developed a procedure to calculate $\hat{\Gamma}({\bf q})$ in the Gaussian approximation using the functional integral representation for the scattering matrix, which is 
\bea
\sigma(\beta)=  \int D \psi {\rm exp}\Big \{ -{1 \over 2\beta} \sum_{\bf q \lambda\lambda'} I^{-1}_{\lambda \lambda'} ({\bf k}) \psi_\lambda ({\bf q}) \psi_{\lambda'} ({-\bf q}) \Big\} \hat{T} {\rm exp}\Big\{ \sum_{\lambda{\bf q}}\psi_\lambda({\bf q}) \delta S^{\lambda} ({\bf q})  \Big\}.
\eea
Here $I^{-1}_{\lambda \lambda'} ({\bf k})$ denotes the inverse matrix elements of $\hat{I}({\bf k})$ indicated in Eq.~(\ref{Eq9}). The functional integration over the complex field variable $ \psi_\lambda({\bf q})=\psi^c_\lambda({\bf q})+i\psi^s_\lambda({\bf q})$  takes the following form, 
\bea
 \int D\psi ... = \Pi_\lambda \int_{-\infty}^{\infty} {d\psi_\lambda(0) \over \sqrt{2\pi {\rm det}[\beta\hat{I}(0)]}} \Pi_{{\bf q} \neq 0}  \nonumber\\ \times \int_{-\infty}^{\infty} {d\psi^c_\lambda({\bf q}) \over \sqrt{\pi {\rm det}[\beta \hat{I}({\bf k})]}} \int_{-\infty}^{\infty} {d\psi^s_\lambda({\bf q}) \over \sqrt{\pi {\rm det}[\beta \hat{I}({\bf k})]}} ...
\eea
Here $\psi^c_\lambda({\bf q})$ and $\psi^s_\lambda({\bf q})$ are the real and imaginary parts of $\psi_\lambda ({\bf q})$, respectively and $\psi^c_\lambda ({-\bf q})=\psi^c_\lambda ({\bf q})$, $\psi^s_\lambda ({-\bf q})=-\psi^s_\lambda ({\bf q})$.

In the Gaussian approximation, Green’s function $\hat{\Gamma}({\bf q})$ satisfies the matrix equation:
\bea
 \hat{\Gamma}({\bf q}) = \hat{M}(\omega)[\hat{1}-\beta\hat{I}({\bf k})\hat{M}(\omega)]^{-1}\,\,,
 \label{Eq29}
\eea
where
\bea
 \hat{M}(\omega)=&\begin{pmatrix}
	\hat{M}_a(\omega) & \hat{O}\\
 	\hat{O}&  \hat{M}_b(\omega) 
\end{pmatrix}, \nonumber\\
=&\begin{pmatrix}
 M_{aX}(\omega)&0&0&0\\
\rm 0&M_{aZ}(\omega)&0&0\\
\rm 0&0& M_{bX}(\omega)&0\\
\rm 0&0&0& M_{bZ}(\omega)\\
\end{pmatrix} ,
\label{Eq30}
\eea
and 
\bea
 M_{\nu X} (\omega) =&  {b_s(y_\nu) \over y_\nu -i\omega\beta},\\
M_{\nu Z} (\omega) =& b'_s(y_\nu)\delta(\omega).
\label{Eq32}
\eea
Here, the Brillouin function $\rm b_s(y_\nu)$ and its derivative $b'_s(y_\nu)$ are given in Ref.~\cite{Bach2019}. 

Using Eqs.(\ref{Eq9})-(\ref{Eq10}), (\ref{Eq29})-(\ref{Eq32}), we evaluate the GFs $\Gamma^{\lambda\lambda'}({\bf q})$ and insert them into Eqs.~(\ref{Eq21a})-(\ref{Eq21}) to obtain the explicit expressions for the GFs in the original spin-lattice coordinate. Performing an analytic continuation, the transverse GF is written by
\bea
G^{xx}_{aa}({\bf q})= {h^2_a b_s(y_a) \over \gamma_a^2} \frac {(\gamma_b-J'\xi_{2{\bf k}}b_s(y_b) - \omega)} { (\omega-\epsilon^{+}_{\bf k})(\omega-\epsilon^{-}_{\bf k})}\,,
\label{Eq33}
\eea
with the k-dependent poles $\epsilon^{\pm}_{\bf k}$ of the GF known as elementary excitation energies or magnons,
\bea
\epsilon^{\pm}_{\bf k} =& {1\over2} \{ 
\gamma_a +\gamma_b - J' \xi_{2\bf k} [ b_s(y_a)+ b_s(y_b) ] \} \nonumber\\
\pm & {1\over 2} \sqrt{ \{\gamma_a-\gamma_b -J'\xi_{2\bf k}[b_s(y_a)-b_s(y_b)]\}^2+ 4C^2 |\xi_{1\bf k}|^2 b_s(y_a) b_s(y_b)}\,. 
\label{Eq34}
\eea
We note that the formula (\ref{Eq34}) better describes the temperature and field dependence of the magnon band structure compared with that obtained by Ref.~\cite{Balatsky2016}. To analyze the magnon spectrum in Eq.~(\ref{Eq34}), we use the MFA result for magnetic moment per site derived from the unitary transformation in Eq.~(\ref{Eq4}), which are
\bea
 m_{\nu z} =& {h_\nu b_s(y_\nu) \over \gamma_\nu}, m_{\nu x} = {\Omega_\nu b_s(y_\nu) \over \gamma_\nu} \,,\nonumber \\
 m_\nu =& \sqrt{m^2_{\nu x}+m^2_{\nu z}} = b_s(y_\nu)\,, \nonumber\\
 m =& (m_a+m_b)/2 \,, \nonumber\\
  y_\nu =& \beta \gamma_\nu . 
  \label{Eq35}
\eea	
We can also obtain various elementary excitations (ferromagnetic, antiferromagnetic, ferromagnetic magnon) in the honeycomb spin-lattice using Eq.~(\ref{Eq34}). In the following parts, we concentrate only on ferromagnetic magnon which has been experimentally observed in the typical monolayer CrI$_3$ and the Van der Waals structures \cite{Huang2017, Cenker2021}.

\section{Ferromagnetic magnon spectra in monolayer honeycomb spin-lattice}

\subsection{Ferromagnetic magnon in the application of transverse field at finite temperature}

 In this part, we examine the magnon spectra in the presence of the spin orientation (SR) effect by virtue of the TF. To emphasize the special role of the transverse field, the longitudinal field is supposedly turned off, $h_0=0$. The ferromagnetic (FM) magnon case corresponds to a homogeneous molecular field where $\gamma_a =\gamma_b \equiv \gamma$, $ h=3(J+2J')m_z$, $\Omega=\Omega_0+3(L+2J') m_x$ and $ C=(J\Omega^2 +Lh^2)/\gamma^2$. Thus, the FM magnon spectrum is given by,
\bea
\epsilon^{\pm}_{\bf k} =  \gamma - [J'\xi_{2\bf k}\mp C|\xi_{\bf 1k}|]b_s(y)\,.
\label{Eq36}
\eea
Since $m=b_s(y)$, the positive spin-wave dispersion relation has two branches only when the net magnetic moment per site $m$ is finite. At a given transverse field, the SR from the out-of-plane to an in-plane direction where $m_z=0,\, m_x=m$ occurs at the SR temperature $\tau_R(\Omega_0)$ derived from the solution of the equation, 
\bea
\rm b_s \Big( {\Omega_0(J+2J') \over \tau_R(J-L)} \Big) = {\Omega_0 \over 3(J-L)}\,,
\label{Eq37}
\eea
with $\rm \beta^{-1}=\tau,\, y=\gamma/\tau$.
Similarly, the Curie temperature is extracted from Eq.~ (\ref{Eq37}) in the limit of zero TF, 
 \bea
 \tau_c = \tau_R(\Omega_0=0) =s(s+1)(J+2J')\,.
 \label{Eq38}
 \eea		     	               

In the presence of TF, the magnetic moment and the spin-wave energy follow different relations in certain temperature regions. Instead of Eq.~(\ref{Eq36}), the energy spectrum of the FM magnon is explicitly expressed in different ranges of temperature which are
 
 i/ Below the SR temperature ($0< \tau < \tau_R(\Omega_0)$)
 \bea
 m_x=&{\Omega_0 \over 3(J-L)}\,, m_z=\sqrt{b^2_s(y)-m_x^2}, \nonumber \\
 \gamma=& 3(J+2J')b_s(y), \,y= \gamma/\tau.
 \label{Eq40}
 \eea 
From that, the spin reorients from the out-of-plane to the in-plane direction at the SR field which is 
\be
\Omega_{0R}(\tau)=3(J-L)b_s(y). 
\label{Eq41}
\ee
The magnon spectrum is then described by
\bea
\epsilon^{\pm}_{\bf k}=b_s(y) \Big\{ 3J+J'(6&-&\xi_{\bf 2k})\nonumber\\
&\pm& |\xi_{\bf 1k}| \Big[ L+{\Omega_0^2 \over 9(J-L)b^2_s(y)} \Big] \Big\},\qquad 
\label{Eq42}
\eea   
 where the zero-momentum magnon mode simplifies to, 
 \bea
 \epsilon_0^{\pm} =3b_s(y) \Big\{ J\pm \Big[ L+{\Omega_0^2 \over 9(J-L)b^2_s(y)}\Big] \Big\}\,.
 \label{Eq43}
 \eea
 At finite temperatures, the gap of the low-energy magnon branch at the center of the BZ closes, $\epsilon^{-}_0=0$, when the transverse field reaches the SR field value in Eq.~(\ref{Eq41}). In this case, the low-energy magnon branch is purely acoustic, which implies $\epsilon_{\bf k}^{-} \rightarrow 0 $ when ${\bf k} \rightarrow 0$.  

 Near the Dirac point K'$\rm \Big( {2\pi\over 3a_0}, -{\rm 2\pi\over 3\sqrt{3}a_0}\Big)$, the magnon energy has a linear wave-vector dependence,
  \bea
  \epsilon^{\pm}({\bf k}) \approx 3b_s(y)(J+3J')\pm\hbar v_m |{\bf k-K'}| \,.
  \eea
  The magnon speed $v_m$ is a function of transverse field and temperature,
  \bea
   v_m = {3b_s(y)a_0 \over 2\hbar} \Big[ L+{\Omega_0^2 \over 9(J-L)b^2_s(y)}\Big]\,.
   \label{Eq45}
  \eea
  
  ii/ Above the SR temperature ($\tau \ge \tau_R$),
  
  The magnetization and the dispersion relation of magnon are correspondingly given by
  
  $ m_z=0$,\, $ m_x=m=b_s(y')$,\, $ y'=\gamma'/\tau$;
  
  $\gamma'= \Omega_0+3(L+2J')b_s(y')$;
 
  $\epsilon^{\pm}_{\bf k}= \Omega_0 +\Big[ 3L+J'(6-\xi_{\bf 2k})\pm J |\xi_{\bf 1k}| \Big]b_s(y')$. 
  
  Near the Dirac point K’, $\epsilon^{\pm}_{\bf k} \approx \Omega_0 + (3L+9J')b_s(y')\pm \hbar v'_m|{\bf k-K'}|$ where the magnon speed is non-explicitly temperature and field dependent, 
  \bea
  v'_m ={3Ja_0b(y') \over 2\hbar}\,.
  \eea
  
  iii/ At zero temperature
  
  The spin reorientation is then interpreted as the quantum phase transition due to the TF. At zero temperature, the SR field equals the critical field $\rm \Omega_{0R}(0)=\Omega_{0C}$. In this case, the Brillouin function reaches a saturated value, $b_s(y)=s$ with the spin moment per site $s$. Consequently, the critical transverse field causing QPT is (see Eq.~(\ref{Eq41}))
 \bea
 \Omega_{0C}=\Omega_{0R}(0)=3s(J-L).
 \label{Eq39}
 \eea
 The spin wave energy behaves differently depending on the transverse field, which is
  
  a) $\Omega_0 \le \Omega_{0C}$,
  \bea
   m_x = {\Omega_0 \over 3(J-L)}\,, m_z=s\sqrt{1-{\Omega_0^2 \over \Omega^2_{0C}}}\,,
  \eea 
   and the FM magnon follows a dispersion relation which is
   \bea
   \epsilon^\pm_{\bf k} = s\Big\{ 3J+(6-\xi_{\bf 2k})J'\pm \Big[ L+{\Omega^2_0 \over 9(J-L)s^2} \Big]  |\xi_{1\bf k}|  \Big\}, \quad
   \eea			
   where the zero-momentum magnon gaps at Brillouin zone center for two branches are  
   \bea
   \epsilon_0^+ =& 3s(J+L)+{\Omega_0^2 \over \Omega_{0C}},\\
   \epsilon_0^- =& \Omega_{0C}\Big(1-{\Omega_0^2 \over \Omega^2_{0C}} \Big).\quad
   \label{Eq50}
   \eea
   It is obviously seen from Eq.~(\ref{Eq50}) that the low-energy branch magnon gap closes at the critical field $\Omega_{0C}$.
   
    b) $\Omega_0 \ge \Omega_{0C}$
    
    The magnetization is simplified to $m_z=0$, $m_x=s$ and the magnon spectrum linearly varies on the TF, which is
    \be
  \epsilon^{\pm}_{\bf k} = \rm  \Omega_0 + s[3L+J'(6-\xi_{2\bf k})\pm J|\xi_{\bf 1k}|]\,.\quad
  \label{Eq51}
    \ee
    \subsection{Numerical results}
 
    \begin{figure}
    \subfloat[\centering][\label{fig:2a}]{{\includegraphics[height=1.8in,width=2.2in]{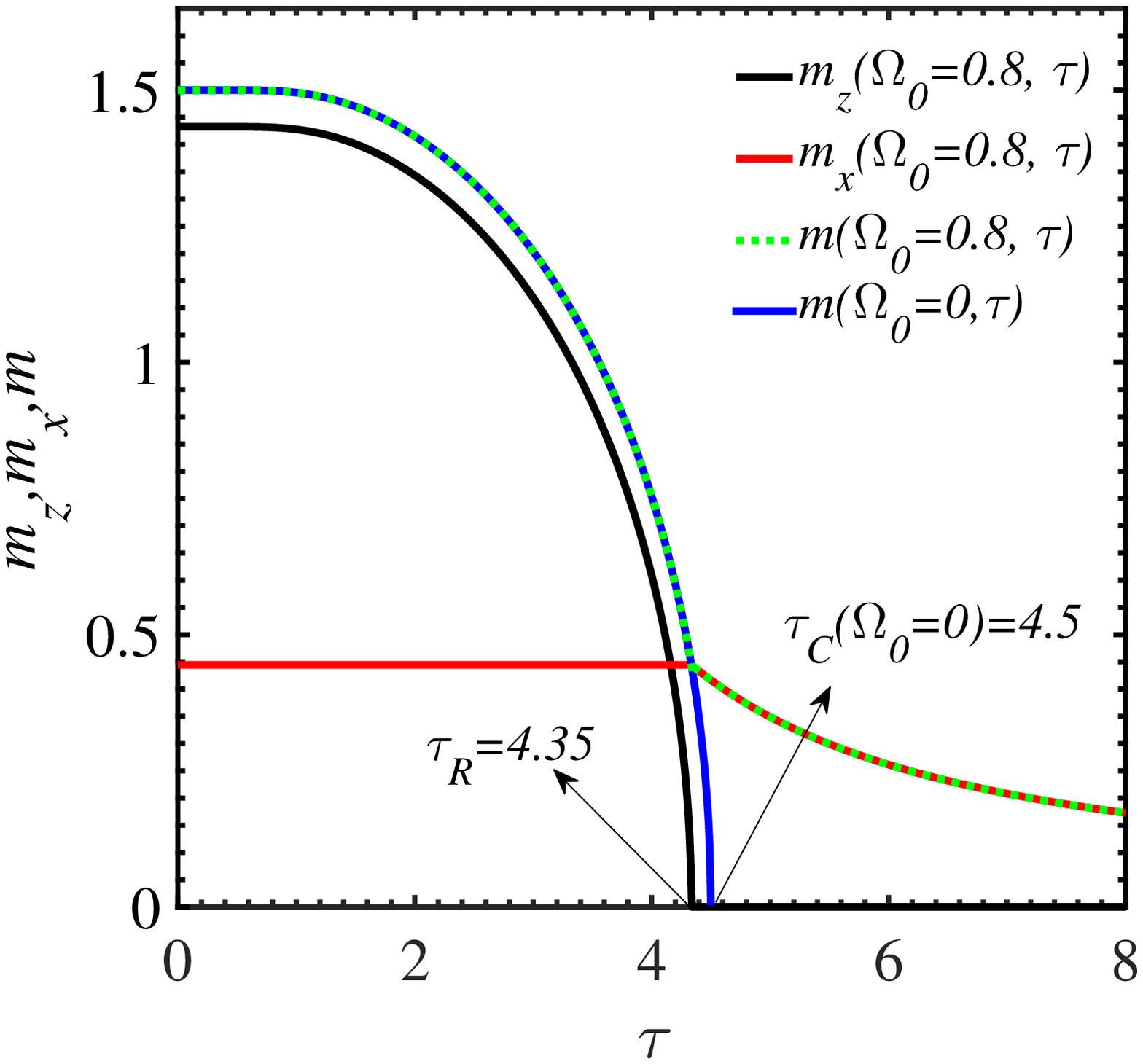} }}%
    \hspace{1cm}
    \subfloat[\centering][\label{fig:2b}]{{\includegraphics[height=1.8in,width=2.2in]{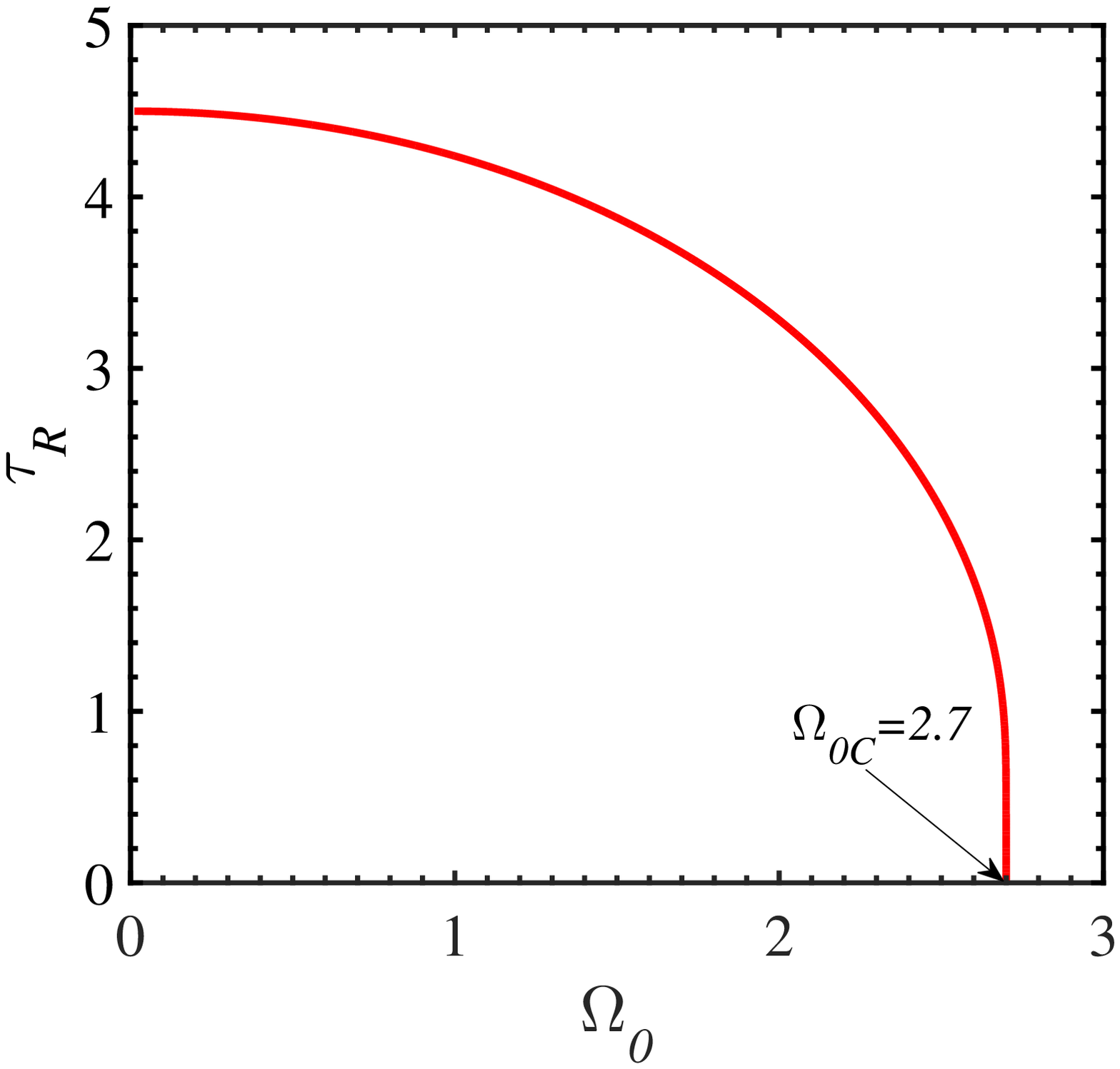} }}%
 \caption{Temperature dependence of the magnetic moment per site m and its component m$\rm_x$, m$\rm_z$ of the FM monolayer honeycomb spin-lattice in the transverse field $\Omega_0=0$ and 0.8 \protect\subref{fig:2a}, and change of spin reorientation temperature with increasing transverse field \protect\subref{fig:2b}. Here s=3/2, L=0.4, J’=0.1.}
 \label{Fig2}
\end{figure}

For numerical calculations, energy-dependent quantities such as FM-magnon energy $\epsilon^{\pm}({\bf k})$, temperature $\tau$; parameters L, J’; field strength $\gamma$, h, $\Omega$ are measured in terms of NN exchange J and lengths are expressed in terms of $a_0$.  
       
    Fig.~\ref{Fig2}\protect\subref{fig:2a} illustrates the MF thermomagnetic behaviors by using Eqs.~(\ref{Eq37})-(\ref{Eq40}), when $\Omega_0$=0 and 0.8, $L=0.4$, $ J’=0.1$, $s=3/2$. The reduction of the spin reorientation temperature $\tau_R$ with increasing transverse field is exhibited in Fig.~\ref{Fig2}\protect\subref{fig:2b}. The Curie temperature $\rm \tau_C=4.5$ is obtained as the spin reorientation temperature when $\Omega_0=0$ according to Eq.~(\ref{Eq38}). At the critical value $\Omega_{0C}=2.7$, the SR temperature becomes zero.
    
    \begin{figure}
    \subfloat[\centering][\label{fig:3a}]{{\includegraphics[height=1.8in,width=2.in]{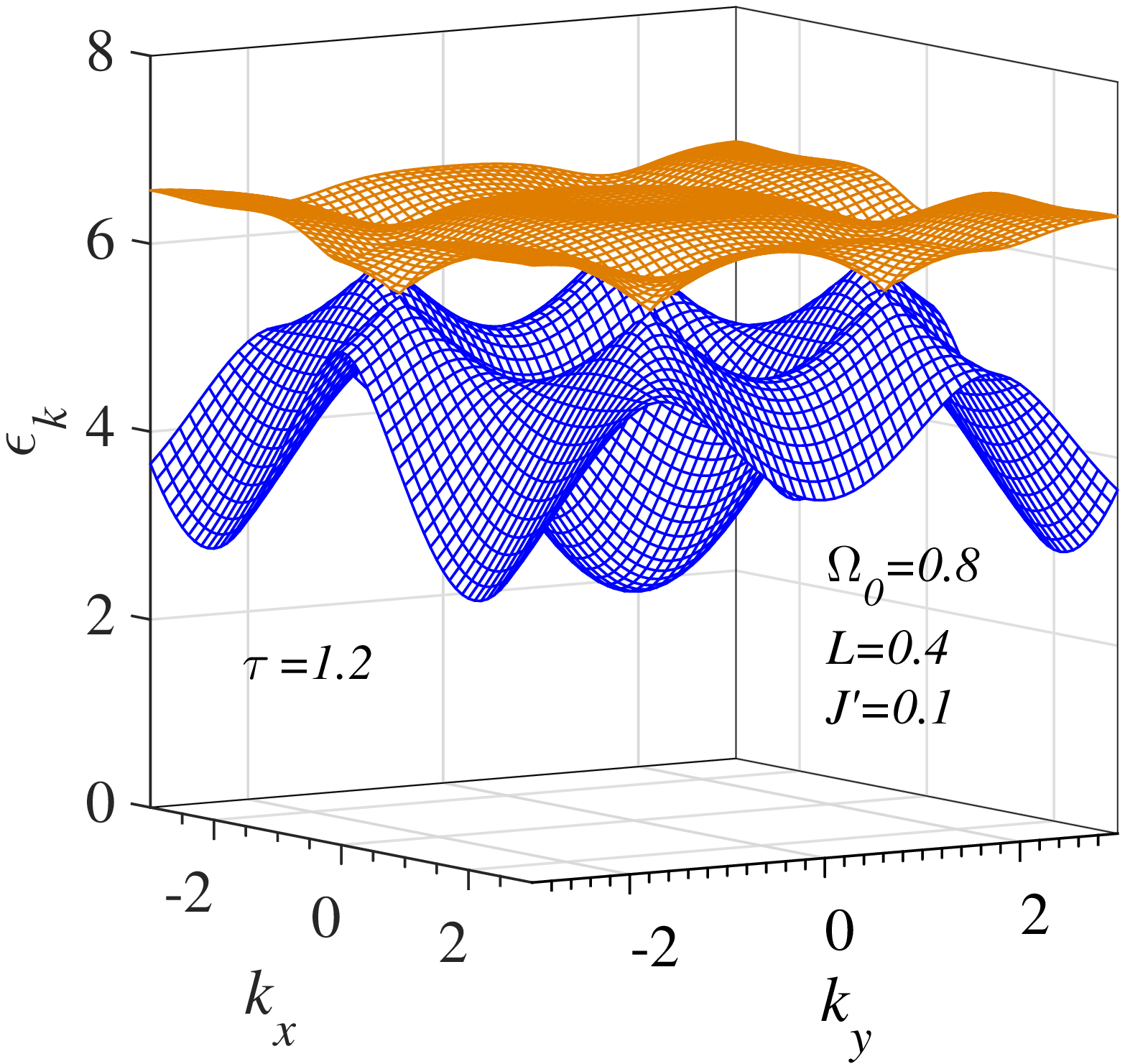}}}%
    \subfloat[\centering][\label{fig:3b}]{{\includegraphics[height=1.8in,width=2.in]{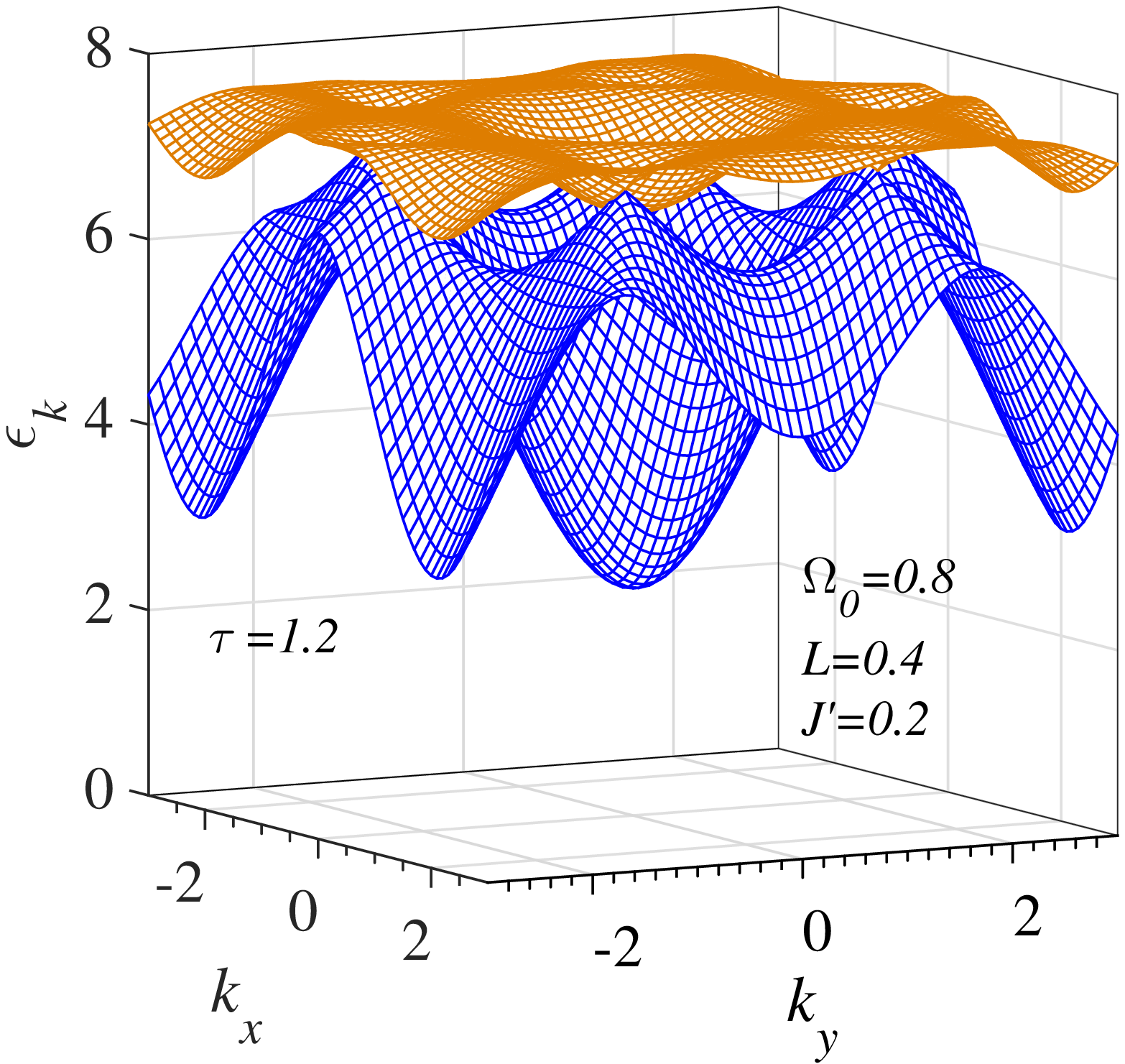} }}%
    \subfloat[\centering][\label{fig:3c}]{{\includegraphics[height=1.8in,width=2.in]{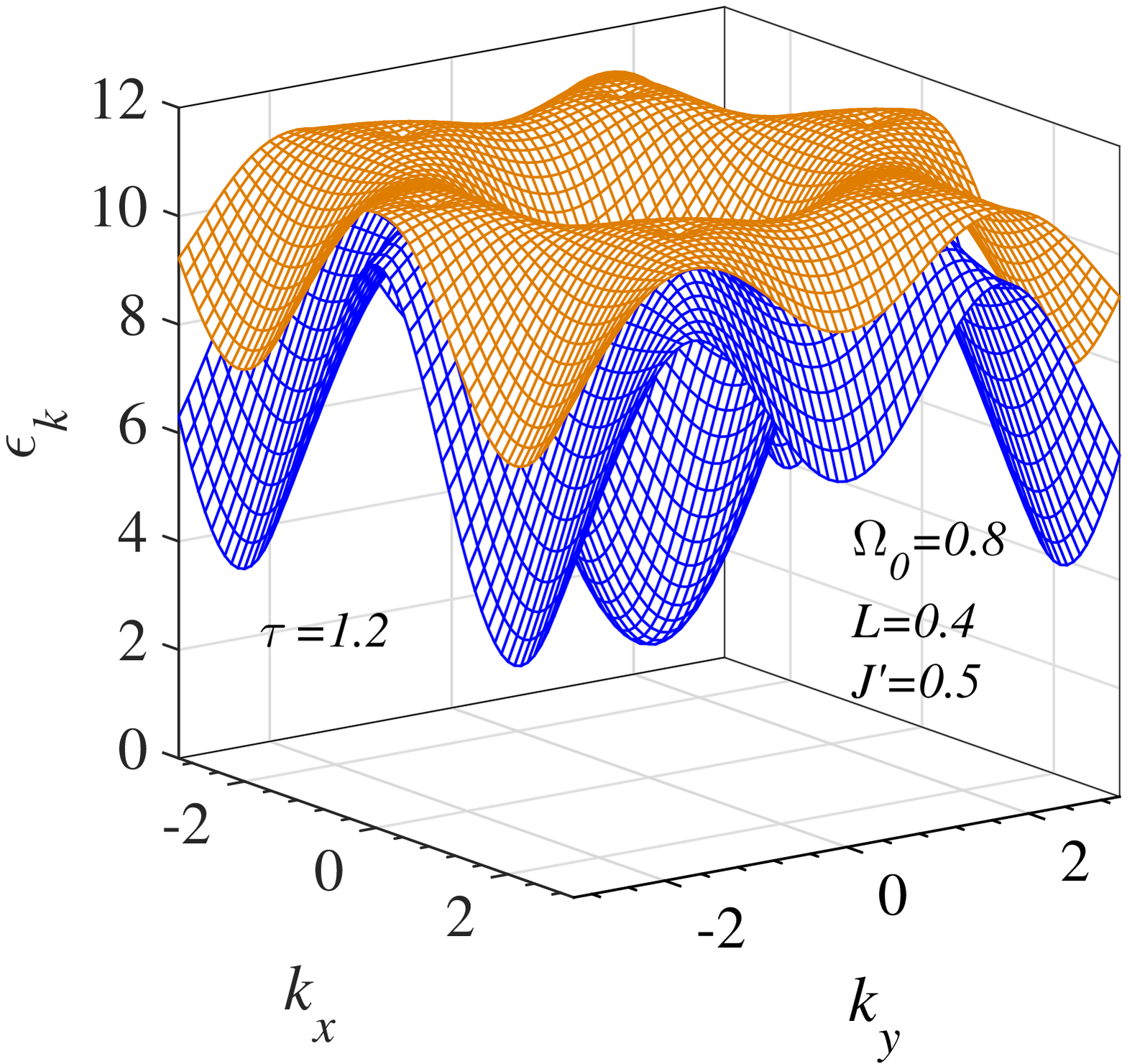} }}%
    \hspace{1cm}
    \subfloat[\centering][\label{fig:3d}]{{\includegraphics[height=1.6in,width=2.in]{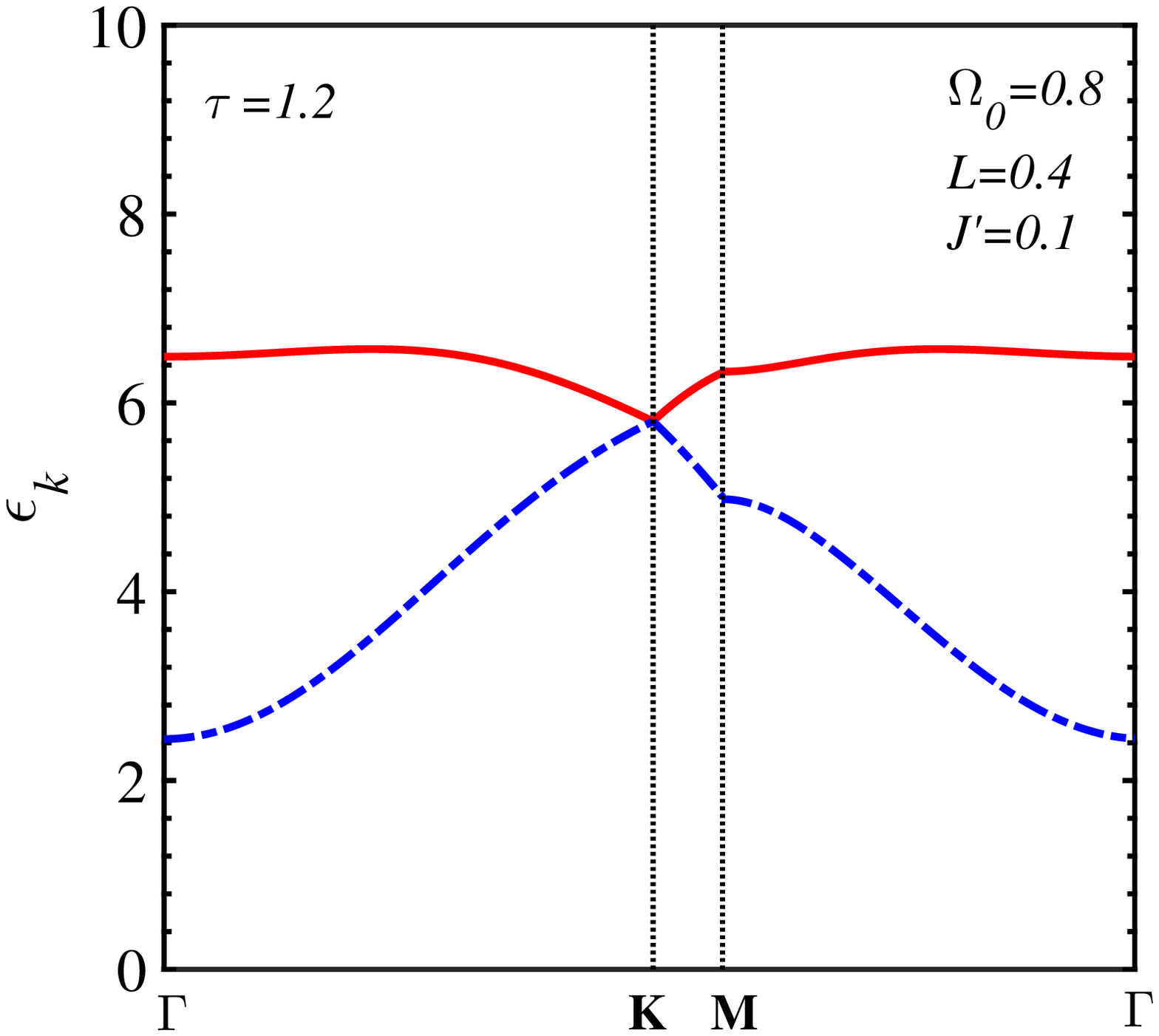} }}%
    \subfloat[\centering][\label{fig:3e}]{{\includegraphics[height=1.6in,width=2.in]{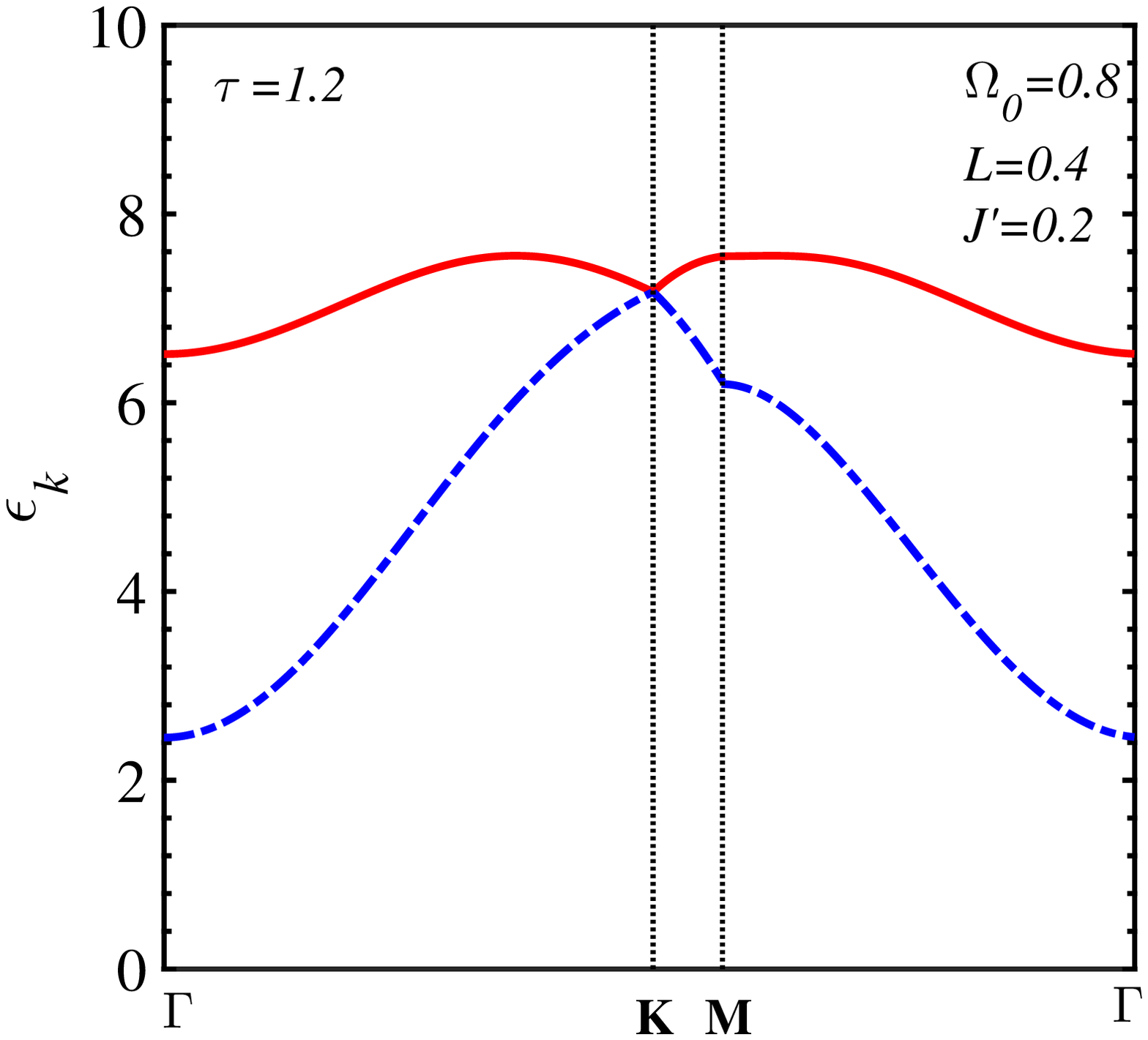} }}%
    \subfloat[\centering][\label{fig:3f}]{{\includegraphics[height=1.6in,width=2.in]{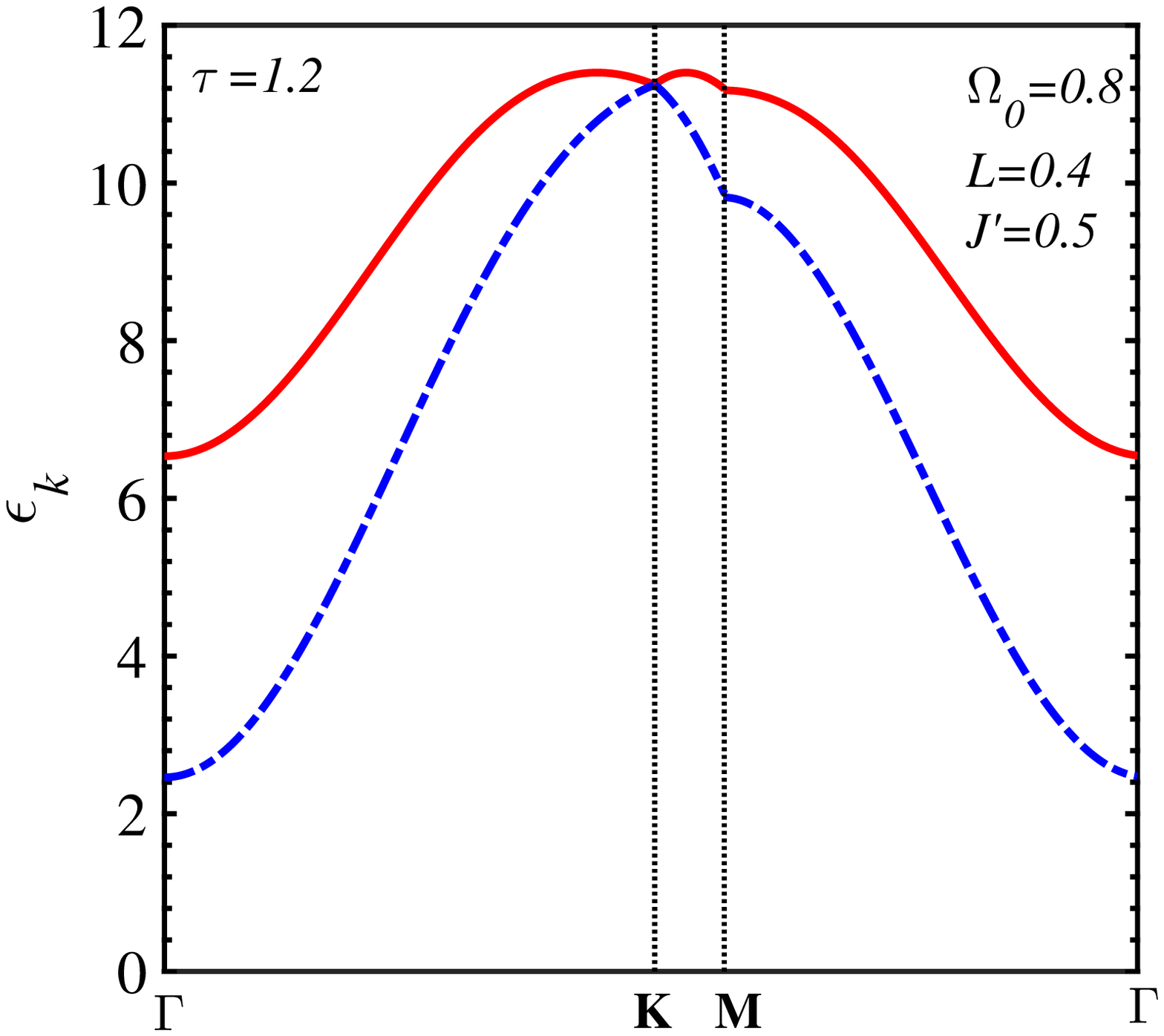} }}%
 \caption{FM magnon spectral branches over the whole Brillouin zone with different NNN interaction $J’$=0.1 \protect\subref{fig:3a}, 0.2 \protect\subref{fig:3b}, 0.5 \protect\subref{fig:3c}. Here transverse field $\Omega_0 = 0.8$, $s=3/2$, $L=0.4$, $\tau=1.2$. The magnon energy spectrum along the $\rm \Gamma MK\Gamma$ line with the same parameters in the \protect\subref{fig:3a}-\protect\subref{fig:3c} cases are illustrated in \protect\subref{fig:3d}-\protect\subref{fig:3f} cases, correspondingly.}
  \label{Fig3}%
\end{figure}

     \begin{figure}
    \subfloat[\centering][\label{fig:4a}]{{\includegraphics[height=1.8in,width=2.2in]{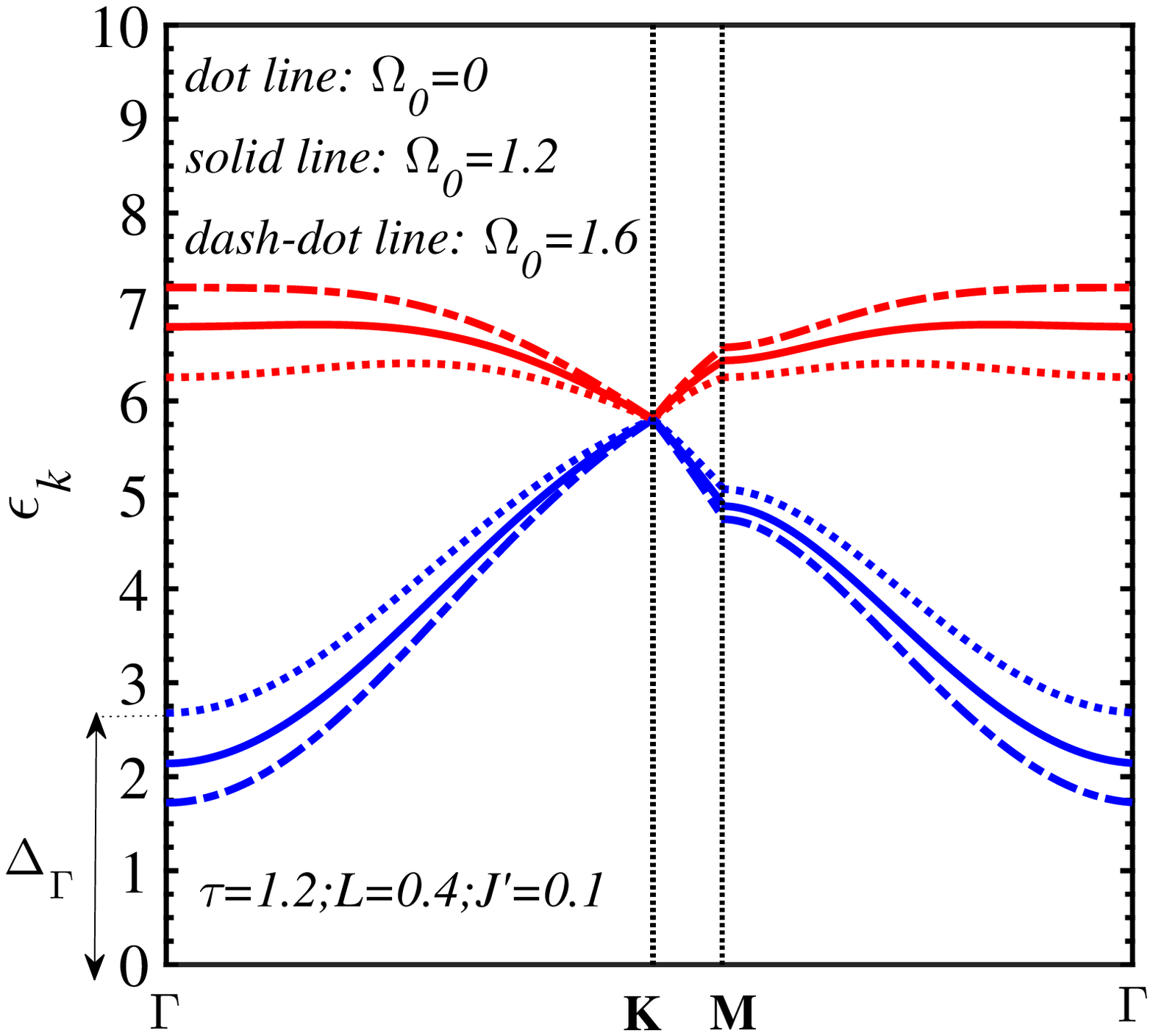} }}%
    \hspace{1cm}
    \subfloat[\centering][\label{fig:4b}]{{\includegraphics[height=1.8in,width=2.2in]{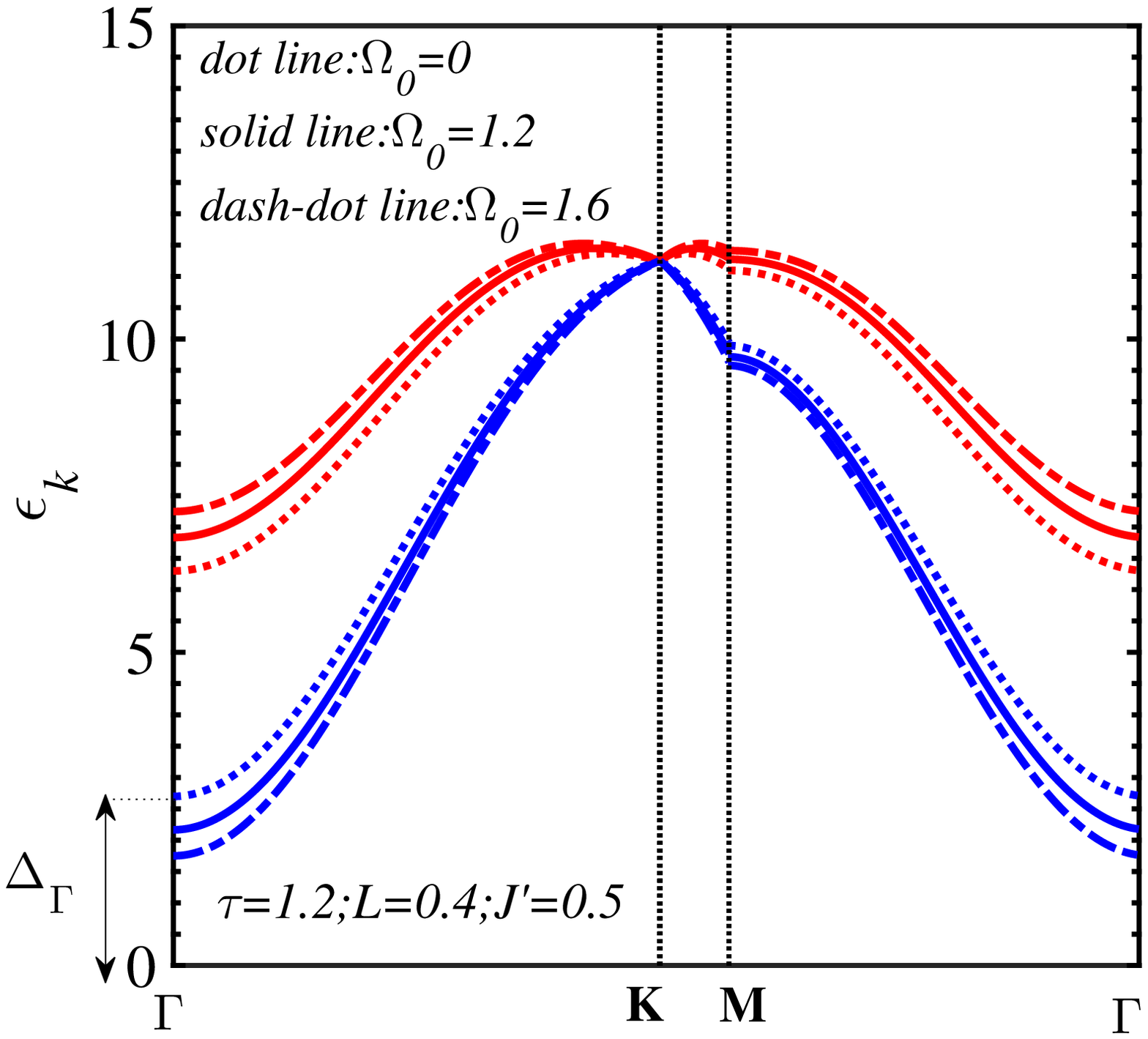} }}%
 \caption{Wave-vector dependence of the magnon branches along the $\rm \Gamma MK\Gamma$ line with increasing transverse field $\Omega_0=$  0, 1.2, 1.6. Parameters are chosen as $s=3/2$, $\tau=$1.2, $L=$ 0.4, and $J’=$ 0.1 \protect\subref{fig:4a}, 0.5 \protect\subref{fig:4b}.}
 \label{Fig4}%
\end{figure}

    Two magnon branches $\epsilon^{\pm}({\bf k})$ are displayed in Fig.~\ref{Fig3}\protect\subref{fig:3a}-\protect\subref{fig:3c} at temperature $\tau=1.2$ (lower than the SR temperature) over the Brillouin zone while the cross-section along the $\rm \Gamma KM\Gamma$ line are presented in Fig.~\ref{Fig3}\protect\subref{fig:3d}-\protect\subref{fig:3f}, respectively. The magnon energy at the Dirac K point increases strongly when increasing the NNN interaction $J’$ from 0.1 to 0.5. The magnon spectra shown 
    in Fig.~\ref{Fig3}\protect\subref{fig:3e}, \protect\subref{fig:3f} are quite similar to the DFT calculation result for the adiabatic magnon energy of CrI$_3$ (see Fig.~8(a) in Ref.~\cite{Wang2020}).
    
    The difference between the magnon branches along the $\rm \Gamma MK\Gamma$ line outside the Dirac K-point increases when the transverse field $\Omega_0$ increases with $J’=0.1$ and 0.5 in the temperature region below $\tau_R$ (see Fig.~\ref{Fig4}). Meanwhile, the upper FM$-$magnon band curvature near the M point changes from upward to downward with increasing $\Omega_0$. This implies that the sign of the effective mass of the upper FM$-$magnon branch is adjustable by the transverse field. As also mentioned in Ref.~\cite{Costa2020}, the energy gap of the low-energy magnon branch at the $\Gamma$ point ($\Delta_\Gamma$ or $\epsilon_0^{-}$) has an anisotropy origin and does not depend on J’ (see Eq.~(\ref{Eq42})). This gap is proportional to the TR field as a function of $\Omega_0^2$ and reduces with increasing TR from 0 to 1.6. 
     \begin{figure}
    \subfloat[\centering][\label{fig:5a}]{{\includegraphics[height=1.8in,width=2.2in]{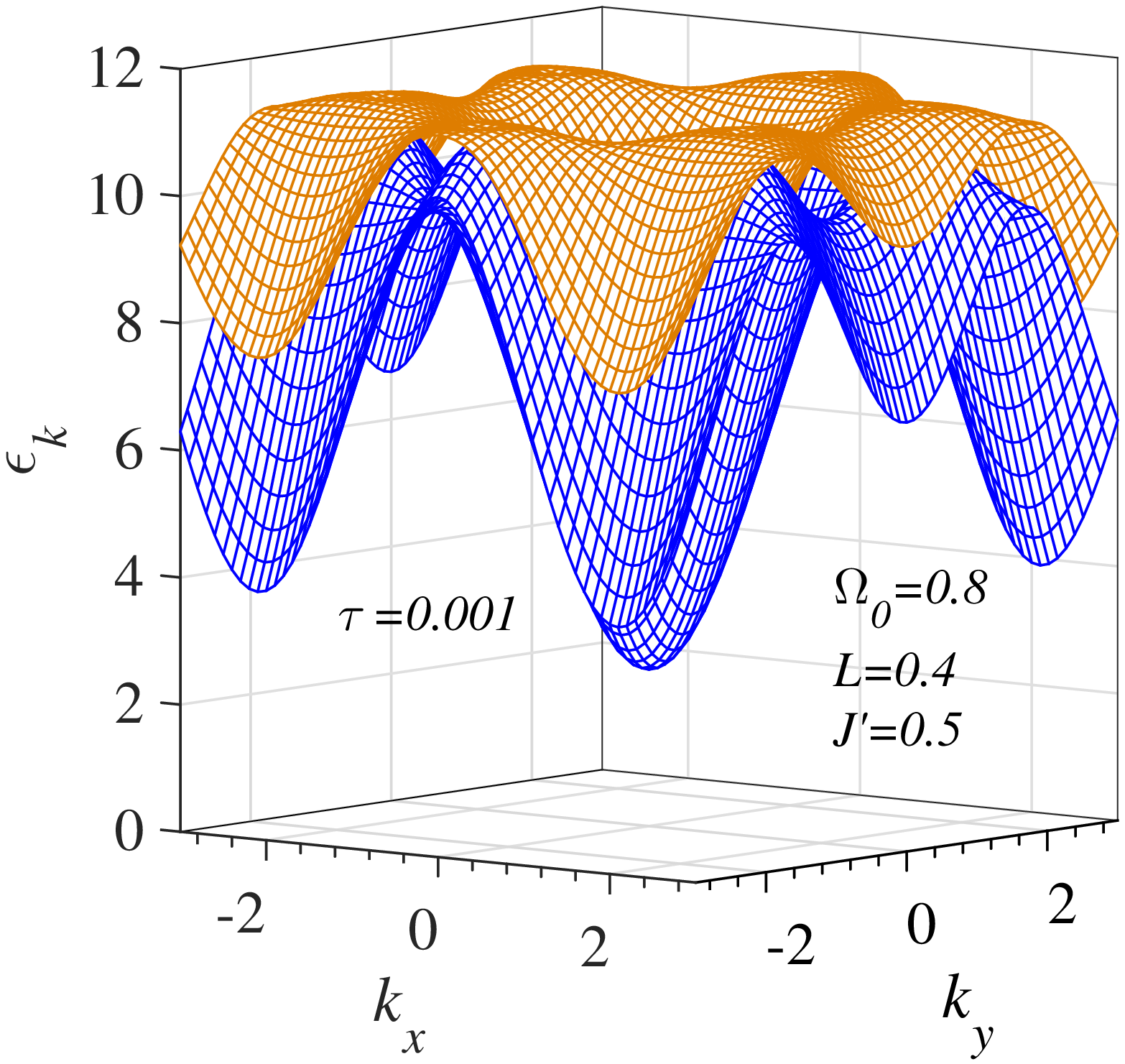} }}%
    \subfloat[\centering][\label{fig:5b}]{{\includegraphics[height=1.8in,width=2.2in]{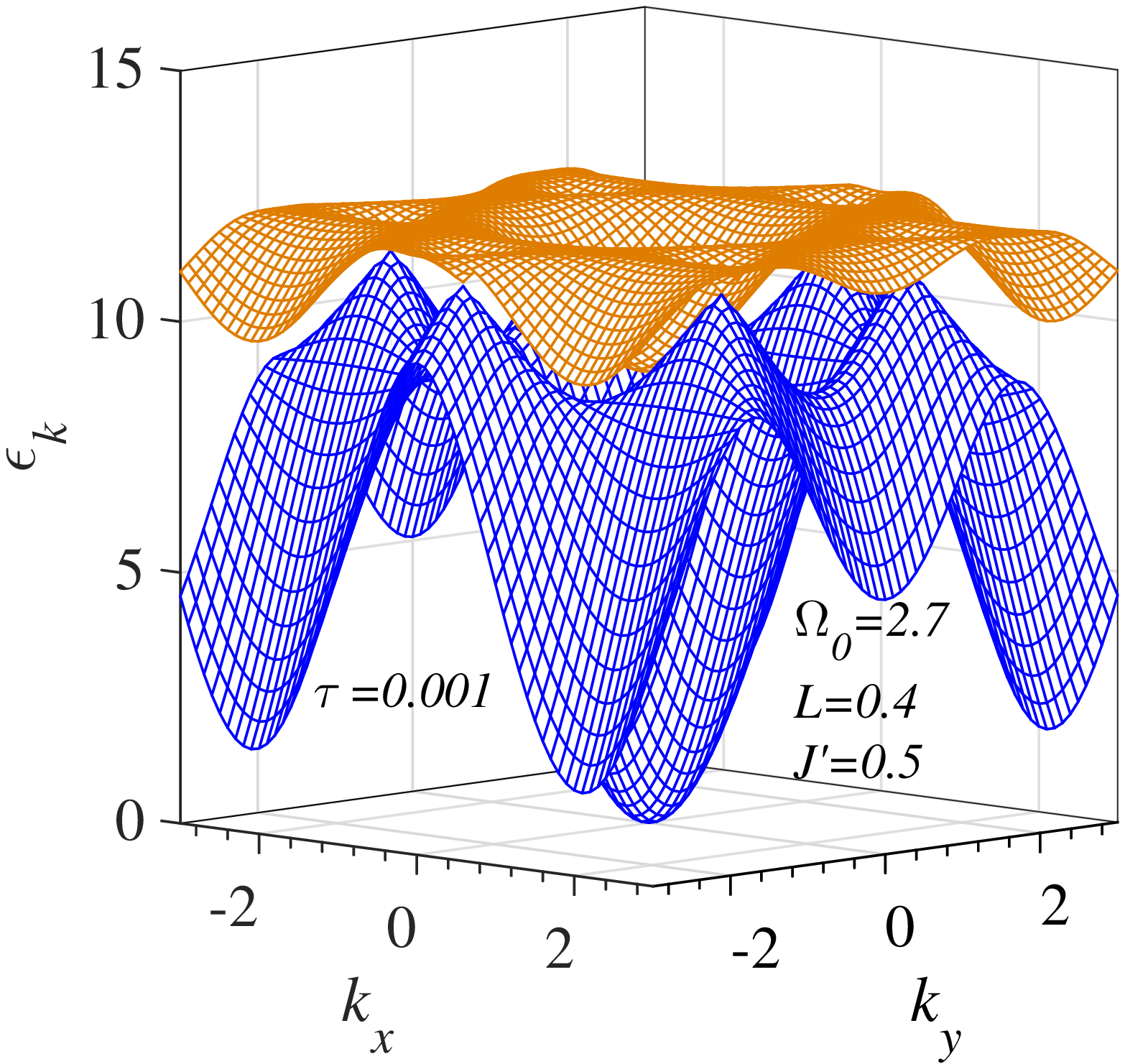} }}%
       \hspace{1cm}
      \subfloat[\centering][\label{fig:5c}]{{\includegraphics[height=1.8in,width=2.2in]{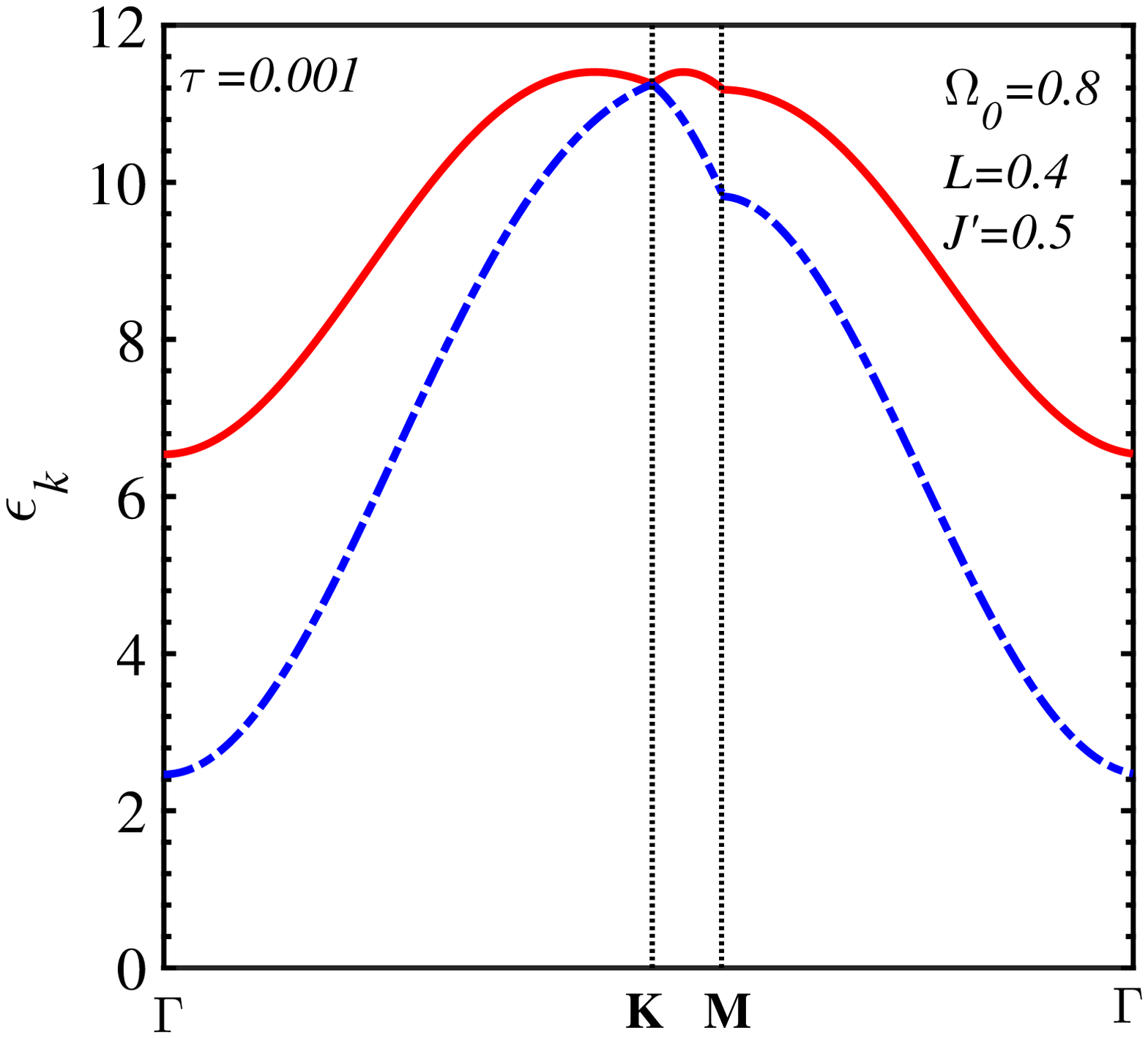} }}%
    \subfloat[\centering][\label{fig:5d}]{{\includegraphics[height=1.8in,width=2.2in]{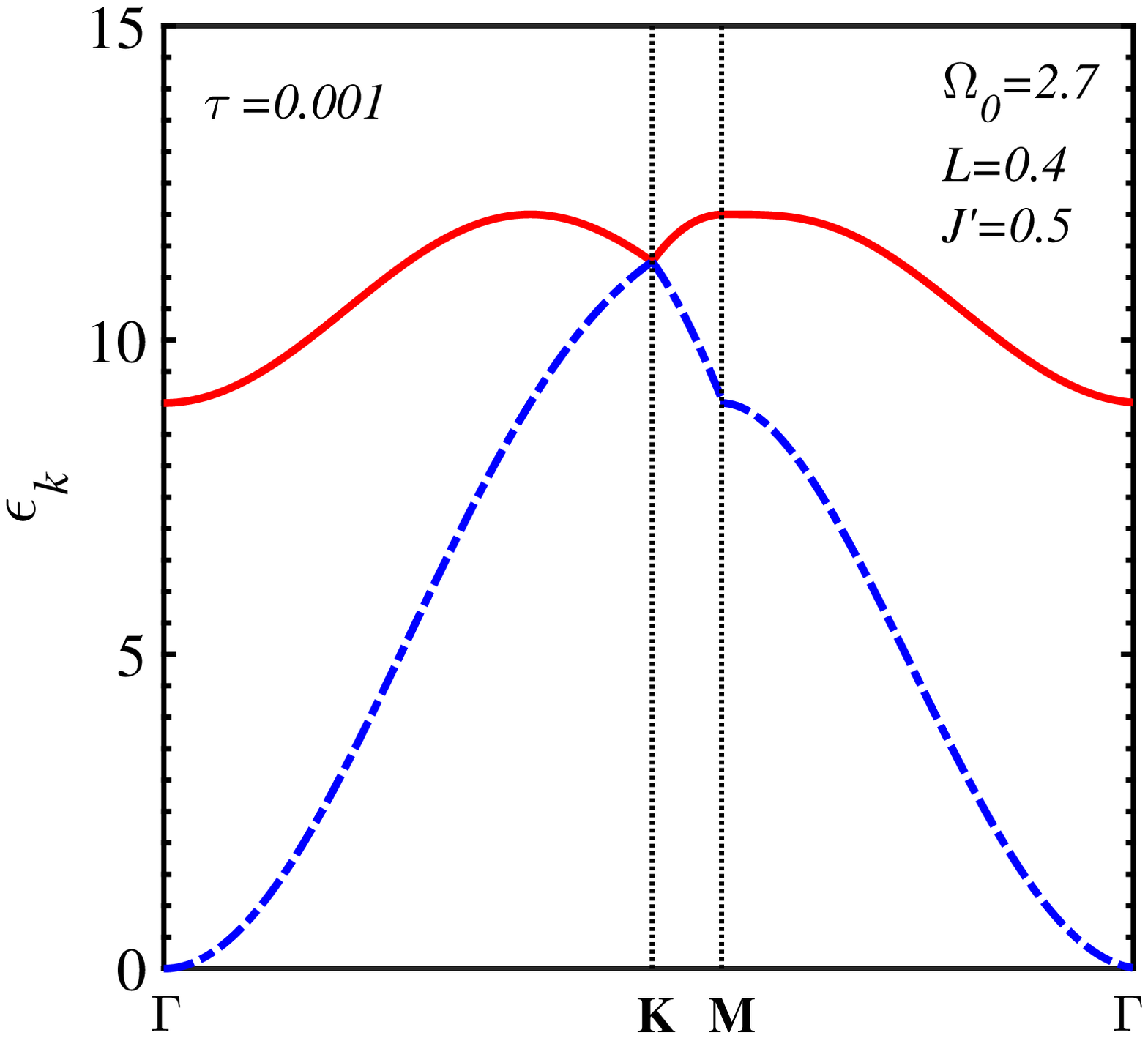} }}%
     \caption{FM$-$magnon branches over the whole Brillouin zone with increasing transverse field $\Omega_0=$ 0.8 \protect\subref{fig:5a} and 2.7 \protect\subref{fig:5b}. Here $s=$3/2, $L=$0.4, $\tau=$ 0.001, $J’=$ 0.5. The magnon energy spectrum along the $\rm \Gamma MK\Gamma$ line with the same parameters in the \protect\subref{fig:5a} and \protect\subref{fig:5b} cases are illustrated in \protect\subref{fig:5c} and \protect\subref{fig:5d}, respectively.}
    \label{Fig5}%
    \end{figure}
    
\begin{figure}
    \subfloat[\centering][\label{fig:6a}]{{\includegraphics[height=1.8in,width=2.2in]{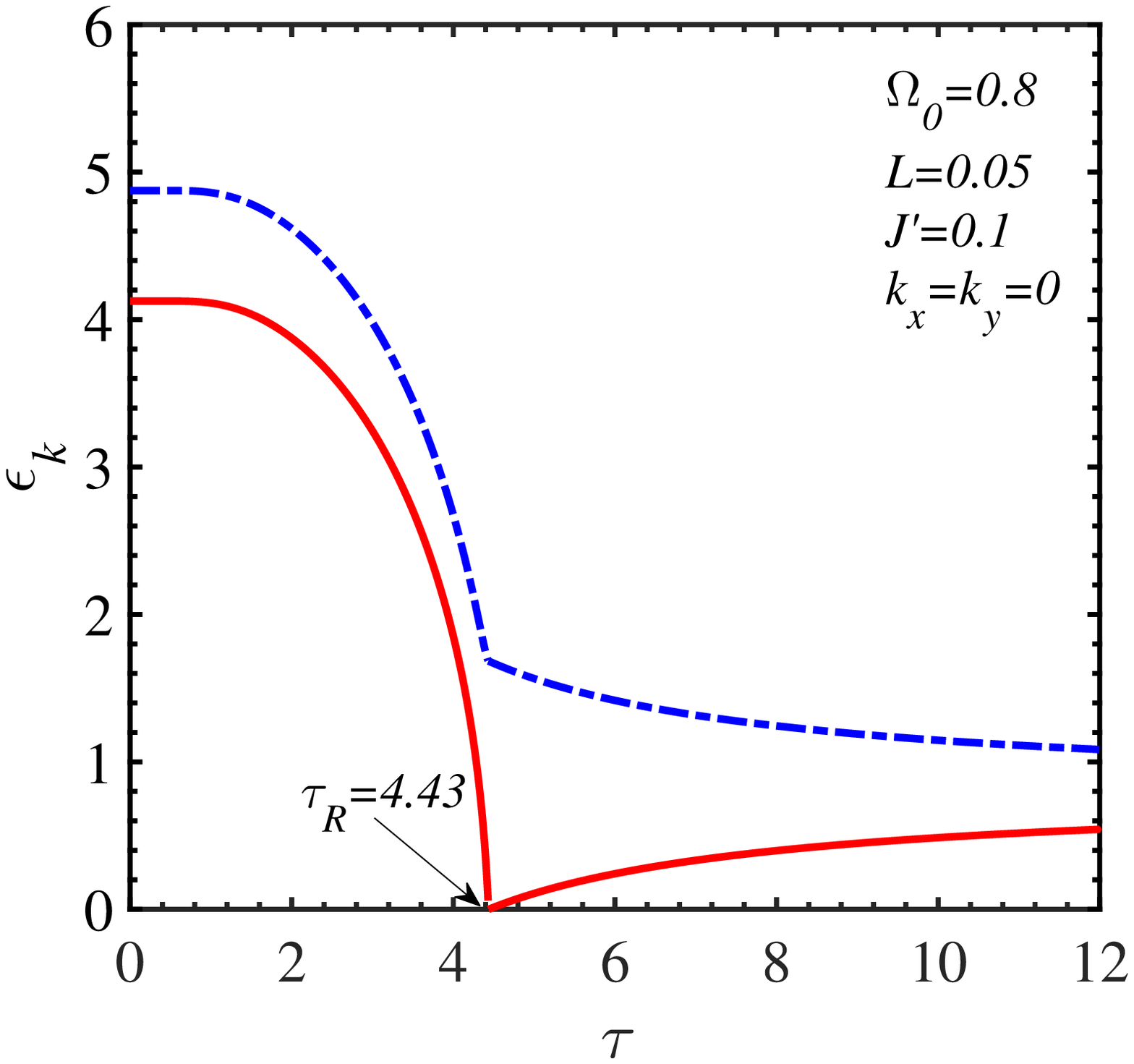} }}%
    \subfloat[\centering][\label{fig:6b}]{{\includegraphics[height=1.8in,width=2.2in]{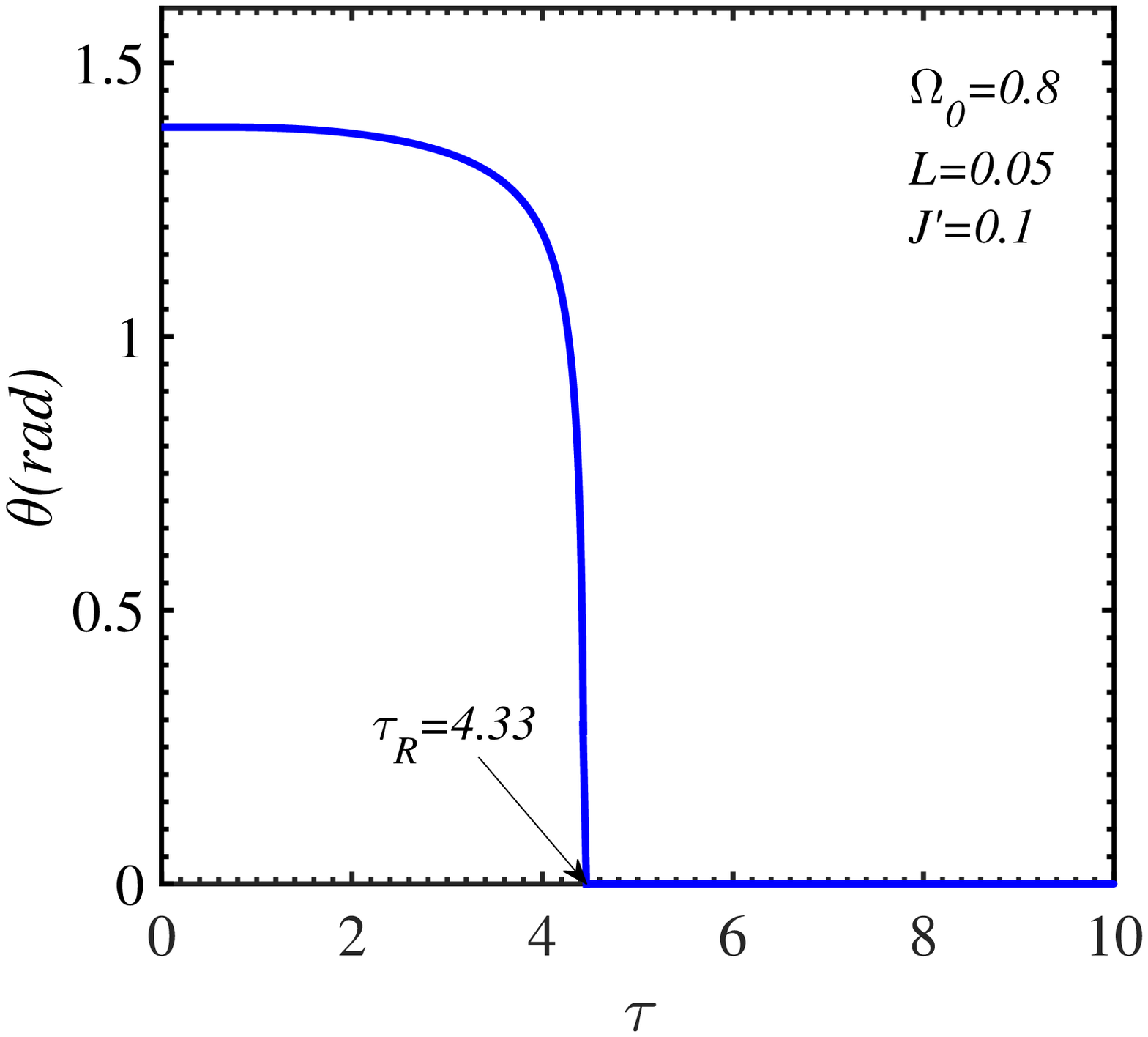} }}%
    \hspace{1cm}
    \subfloat[\centering][\label{fig:6c}]{{\includegraphics[height=1.8in,width=2.2in]{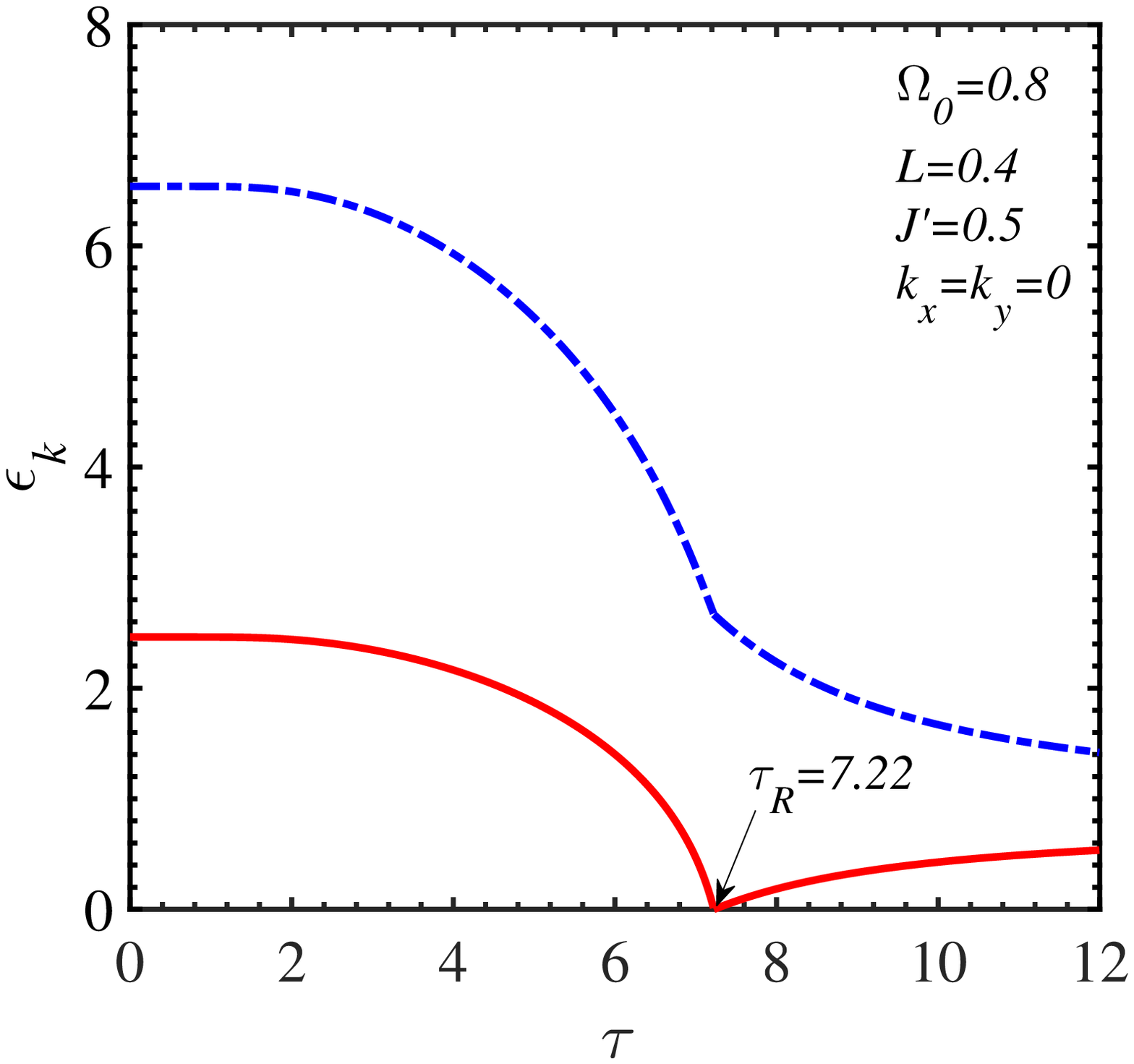} }}%
    \subfloat[\centering][\label{fig:6d}]{{\includegraphics[height=1.8in,width=2.2in]{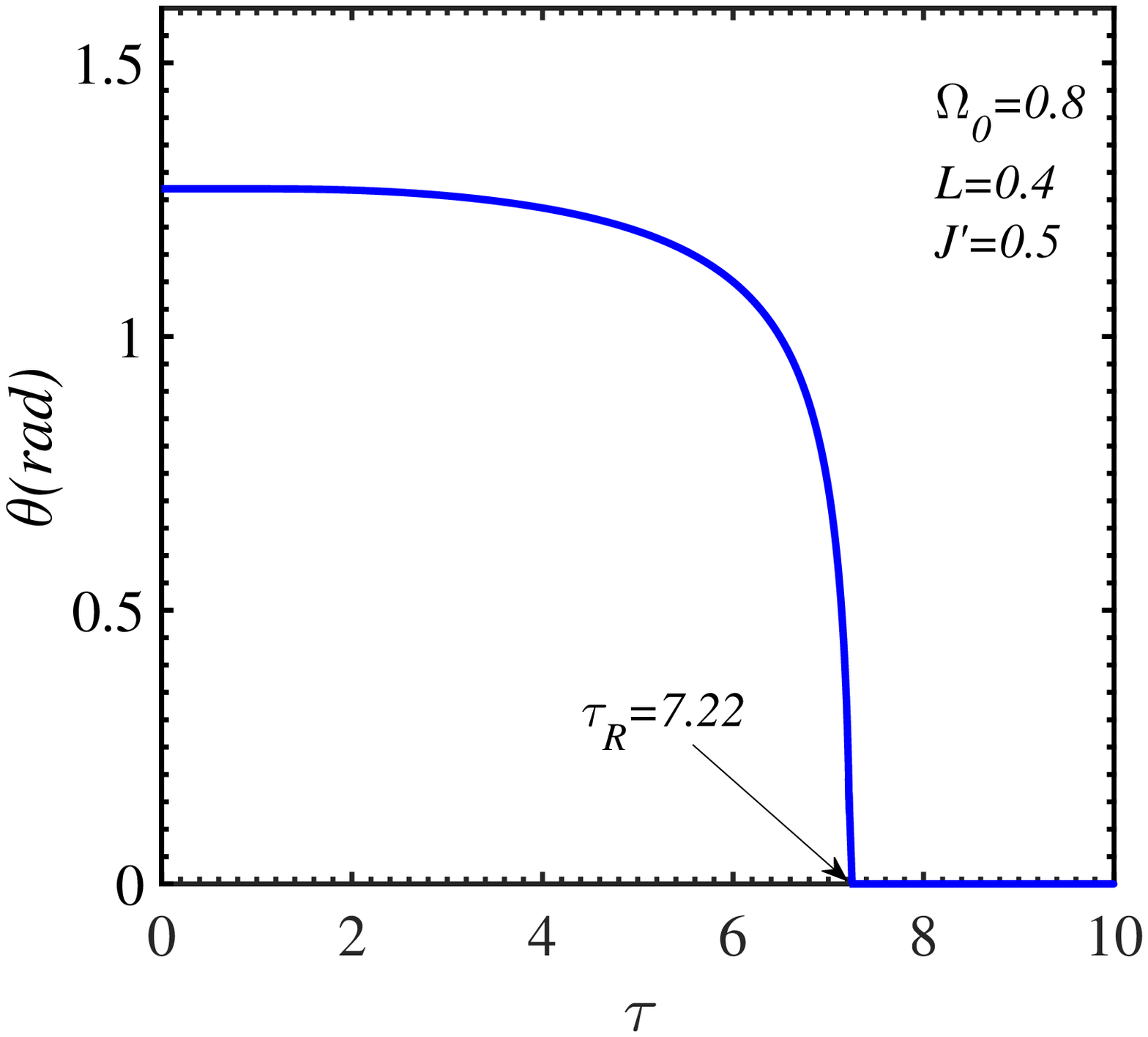} }}%
 \caption{The zero-momentum (${\bf k}=0$) spin-wave energy and the decline angle $\theta$ of magnetic-moment vector versus temperature in the transverse field $\Omega_0=$ 0.8. Other parameters are chosen as $L=$ 0.05, $J’=$ 0.1 for \protect\subref{fig:6a} and \protect\subref{fig:6c}; $L=$ 0.4, $J’=$ 0.5 for \protect\subref{fig:6b} and \protect\subref{fig:6d}. Arrows indicate the SR temperatures. Here the spin value is taken $s=$ 3/2.}
  \label{Fig6}%
\end{figure}

     Fig.~\ref{Fig5} exhibits the magnon branches along the $\rm \Gamma KM\Gamma$ line at very low temperature $\tau=0.001$. At the center of the Brillouin zone, or the $\rm \Gamma$ point, the upper FM$-$magnon branch bottom shifts to the higher value with increasing the transverse field, and the gap of the low-energy FM$-$magnon branch closes when the transverse field approaches the spin reorientation field $\Omega_{0R}=2.7$. The temperature dependence of the zero-momentum FM$-$magnon spectra $\epsilon_{\bf k}({\bf k=0})$ in Fig.~\ref{Fig6}\protect\subref{fig:6a}, \protect\subref{fig:6c} indicates that two branches of these spectra change discontinuously at the SR temperature corresponding to the step-change of the incline angle of the spin direction above the single-layer spin plane as seen in Fig. \ref{Fig6}\protect\subref{fig:6b}, \protect\subref{fig:6d}. For the case of Fig.~\ref{Fig6}\protect\subref{fig:6a} ($L=0.05$, $J'=0.1$), our model is analogous to the Ising model where the zero-momentum magnon energies of two branches are weak temperature-dependent in the temperature region above $\rm \tau_R$. This realization was also experimentally observed in Ref.~\cite{He2018}.

    \begin{figure}
    \centering
    \subfloat[\centering][\label{fig:7a}]{{\includegraphics[height=1.8in,width=2.in]{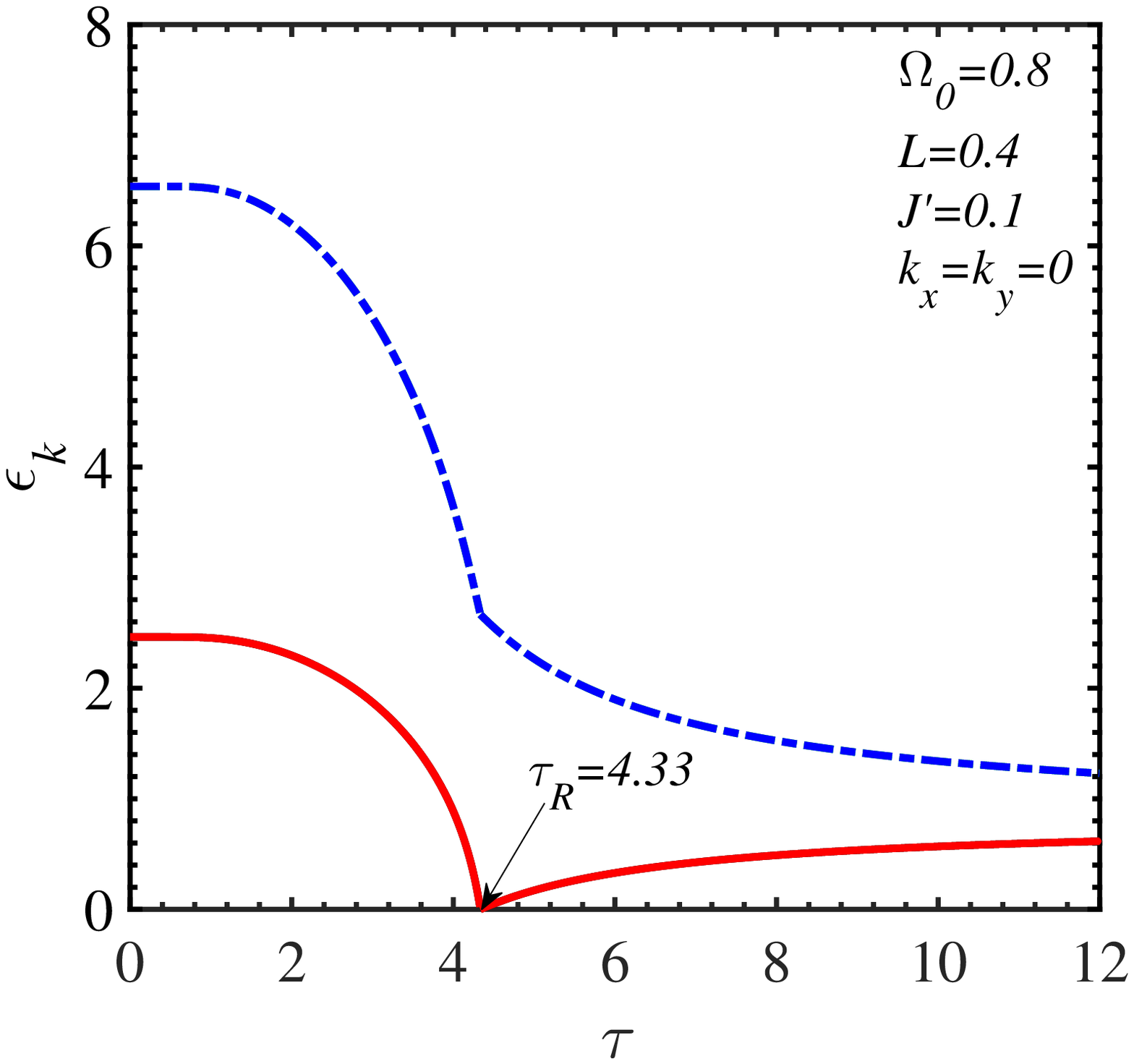} }}%
    \subfloat[\centering][\label{fig:7b}]{{\includegraphics[height=1.8in,width=2.in]{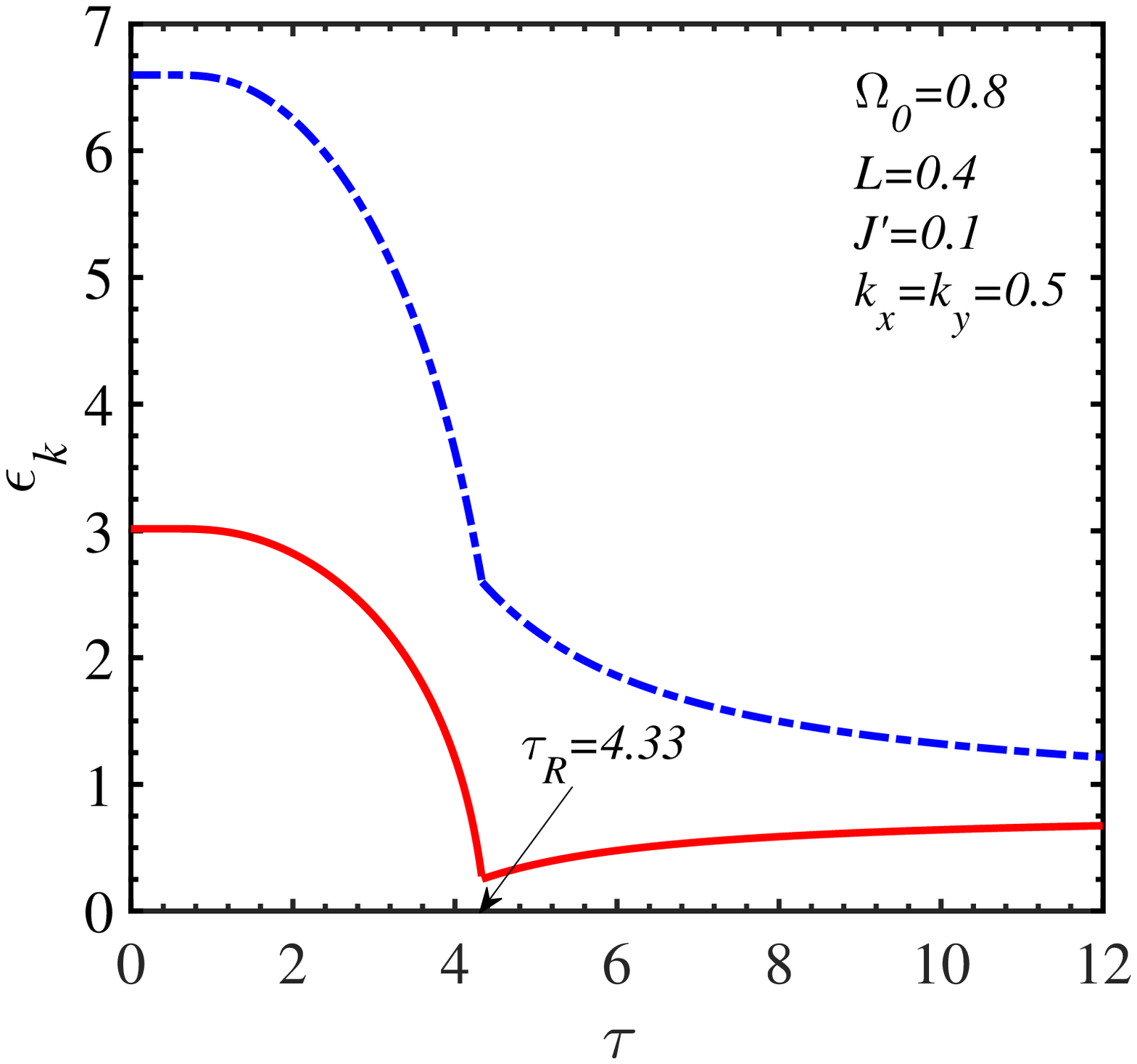} }}%
     \subfloat[\centering][\label{fig:7c}]{{\includegraphics[height=1.8in,width=2.in]{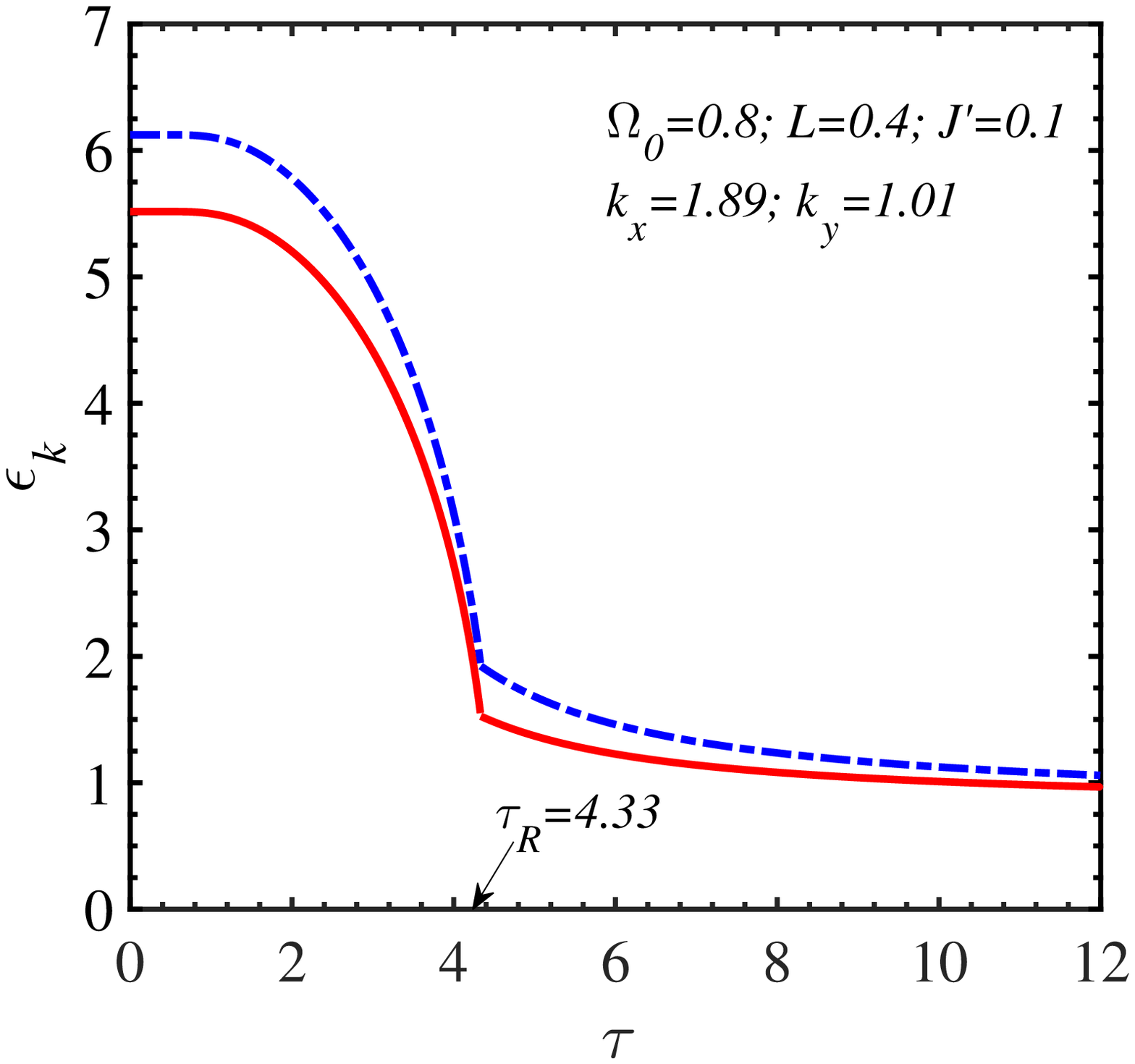} }}%
 \caption{The discontinuous change of the FM magnon spectrum at the SR temperature when $k_x=k_y=0$ \protect\subref{fig:7a}; $k_x=k_y=0.5$ \protect\subref{fig:7b}; $k_x=1.89,\,k_y=1.01$ \protect\subref{fig:7c}. Here we choose $\Omega_0 = 0.8$, $L=0.4$, $J’=0.1$, $s=3/2$.}
 \label{Fig7}%
\end{figure}

  \begin{figure}
    \subfloat[\centering][\label{fig:8a}]{{\includegraphics[height=1.8in,width=2.in]{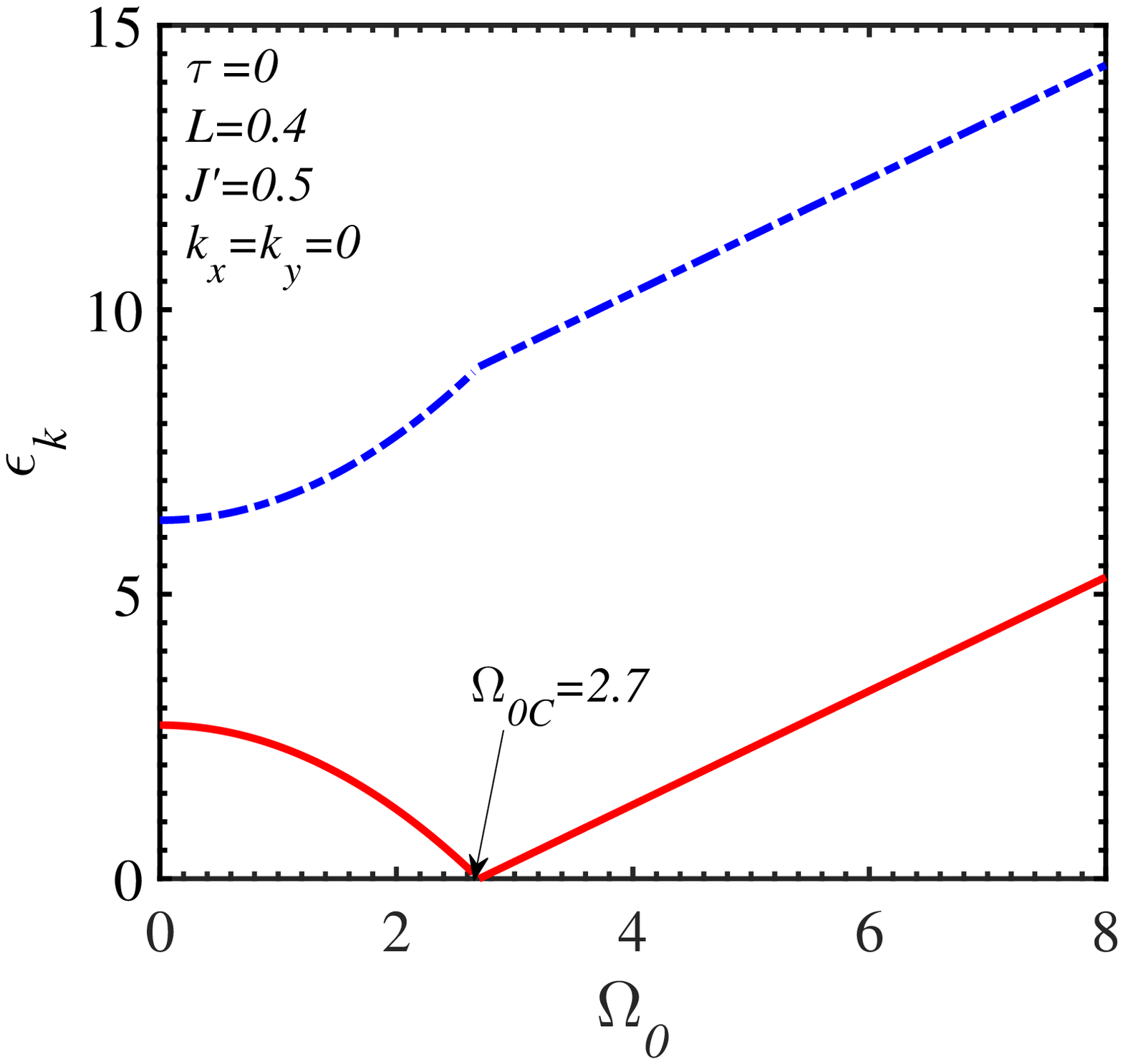} }}%
      \subfloat[\centering][\label{fig:8b}]{{\includegraphics[height=1.8in,width=2.in]{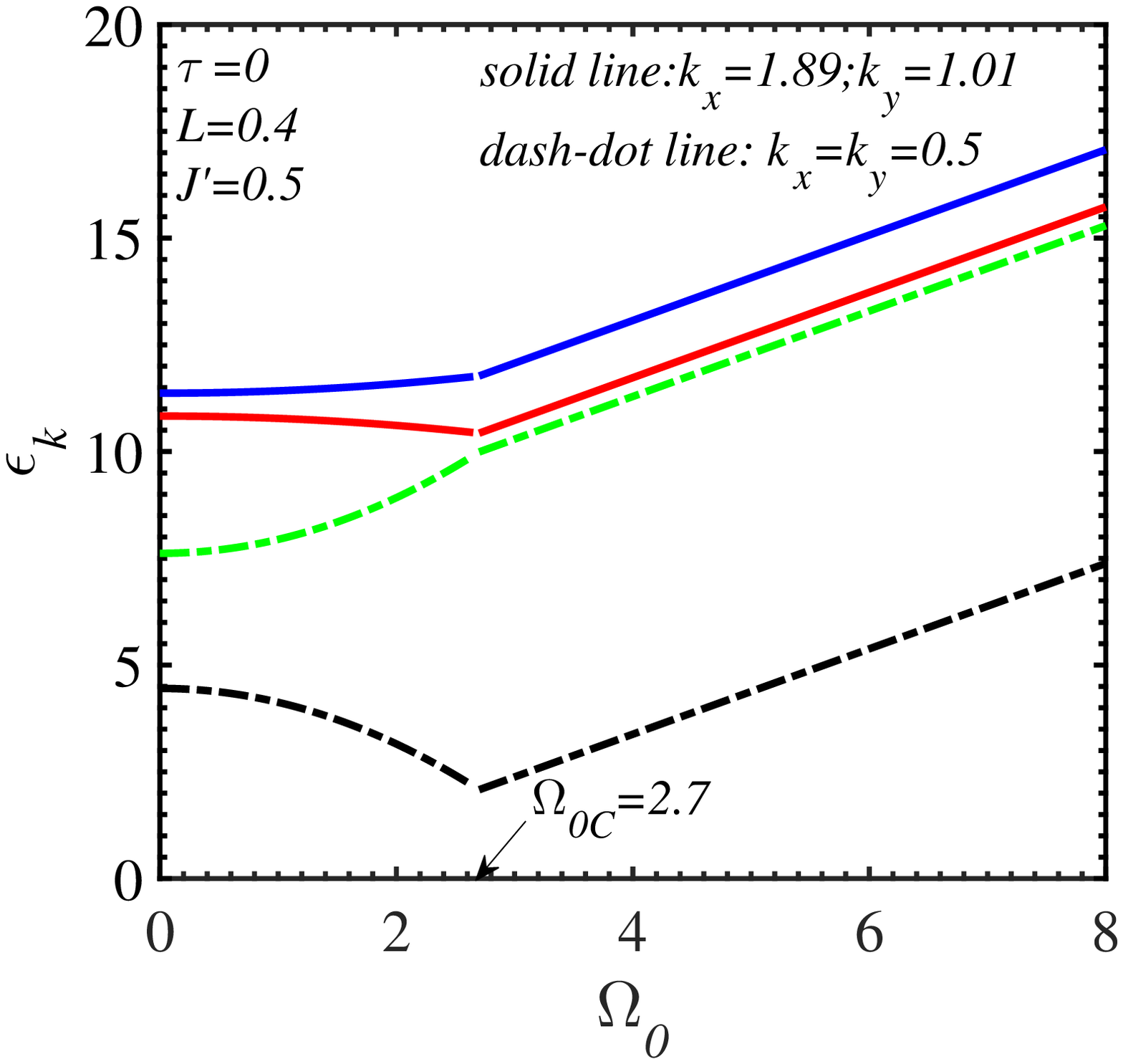} }}%
    \subfloat[\centering][\label{fig:8c}]{{\includegraphics[height=1.8in,width=2.in]{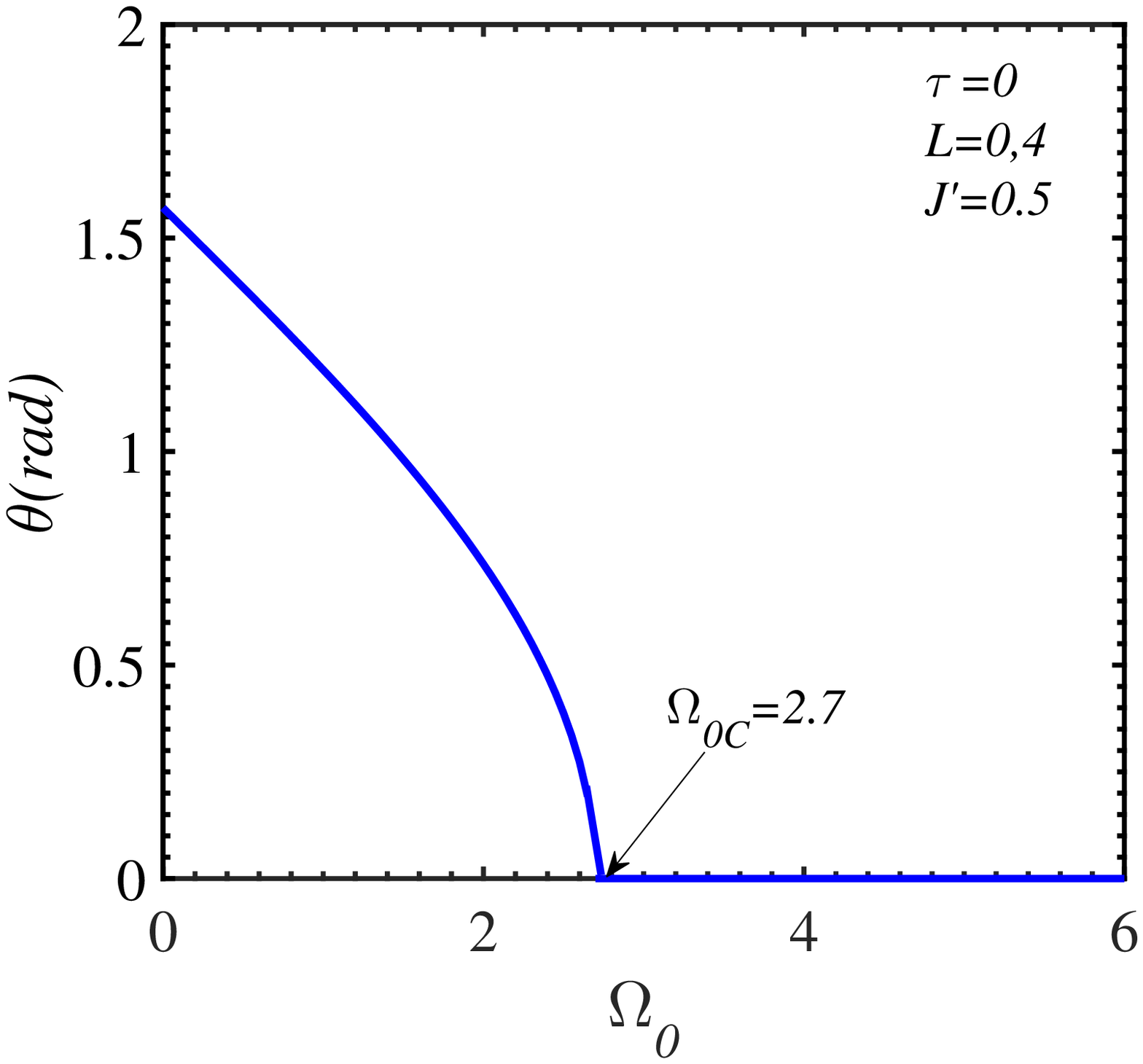} }}%
 \caption{The magnon energy modes at different points of the Brillouin zone where $k_x=k_y=0$ \protect\subref{fig:8a}; $k_x=k_y=0.5$ and $k_x=1.89,\,k_y=1.01$ \protect\subref{fig:8b}, and the inclined angle of the magnetic moment versus the transverse field \protect\subref{fig:8c}. Here we choose $L=0.4$, $J’=0.5$, $s=3/2$.}
  \label{Fig8}%
\end{figure}

    The influence of temperature on the behavior of the magnon energy is unveiled in Fig.~\ref{Fig7} where a disrupt change of $\epsilon_{\bf k}$ occurs at the SR temperature. At a certain transverse field, the difference between the two magnon branches depends on the intrinsic parameters $L$, $J’$, $s$, and the wave vector. For the same set of parameters, $\Omega_0 = 0.8$, $L=0.4$, $J’=0.1$, $s=3/2$, this difference is more pronounced for the wave vector at or near the center of the Brillouin zone ($k_x=k_y=0.5$ in Fig.~\ref{Fig7}\protect\subref{fig:7b}) than for the $\bf k$ value near the Dirac points ($k_x=1.89; \,k_y=1.01$ in Fig.~\ref{Fig7}\protect\subref{fig:7c})). Moreover, it is also indicated that a possible temperature gap opens for the non-zero-momentum low-energy magnon branch. In other words, this branch can not vanish in a wide range of temperatures.
  
    At zero temperature $\tau=0$ where the SR transition is considered as the QPT with the tuning parameter $\Omega_0$, we are interested in the energy variation of several magnon modes with various transverse field $\Omega_0$. The gap between two branches of the zero-momentum FM$-$magnon energy shown in Fig.~\ref{Fig8}\protect\subref{fig:8a} increases proportionally to $\Omega_0^2$ when $\Omega_0 \le \Omega_{0C}$ and reaches the maximum value of 6$s$ for $\Omega_0 \ge \rm \Omega_{0C}$. It is emphasized that the zero-momentum magnon branch cuts the horizontal axis in Fig.~\ref{Fig8}\protect\subref{fig:8a} at the critical transverse field $\rm \Omega_{0C}$. The low-energy magnon branch mode at other points of Brillouin zone ($\bf{k} \neq 0$) including points near the Dirac one ($k_x=1.89;\,k_y=1.01$ in Fig.~\ref{Fig8}\protect\subref{fig:8b} is always finite for any TF value. The gap between two magnon branches remains constant when the TF intensity is larger than the critical tuning parameter $\Omega_{0C}$ due to the linear dependence on the TF displayed in Eq.~(\ref{Eq51}). We also present in Fig.~\ref{Fig8}\protect\subref{fig:8c} the change of the inclined angle of the spin direction $\theta$ with different TF $\Omega_0$. Both the inclined angle and the SR temperature disappear at the critical field $\Omega_{0C}=2.7$ (see also Fig.~\ref{Fig2}\protect\subref{fig:2b}).
  
    \subsection{Estimation for CrI$_3$ monolayer}
  
\begin{figure}
\subfloat[\centering][\label{fig:9a}]{{\includegraphics[height=1.8in,width=2.in]{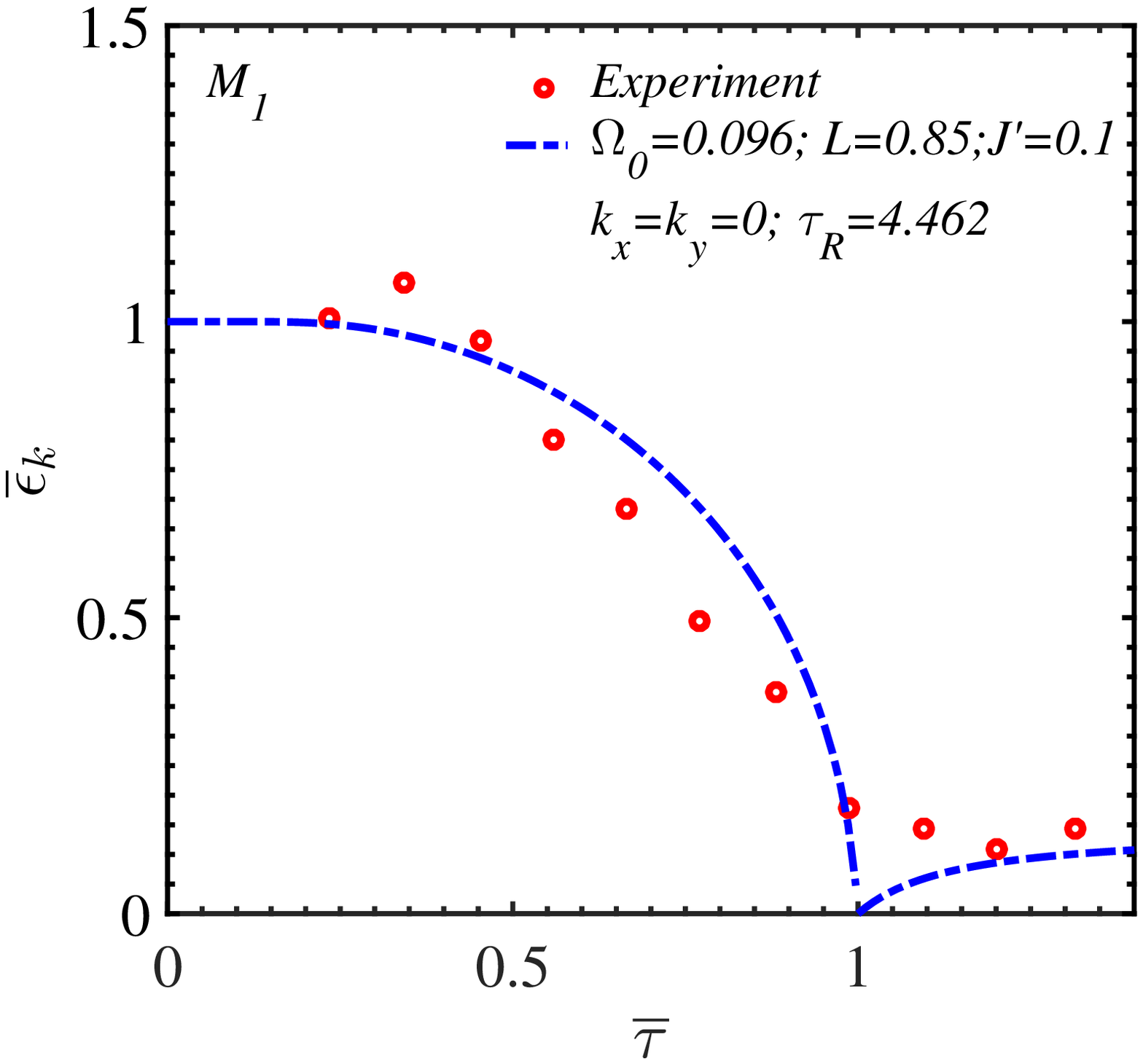} }}%
\hspace{1cm}
\subfloat[\centering][\label{fig:9b}]{{\includegraphics[height=1.8in,width=2.in]{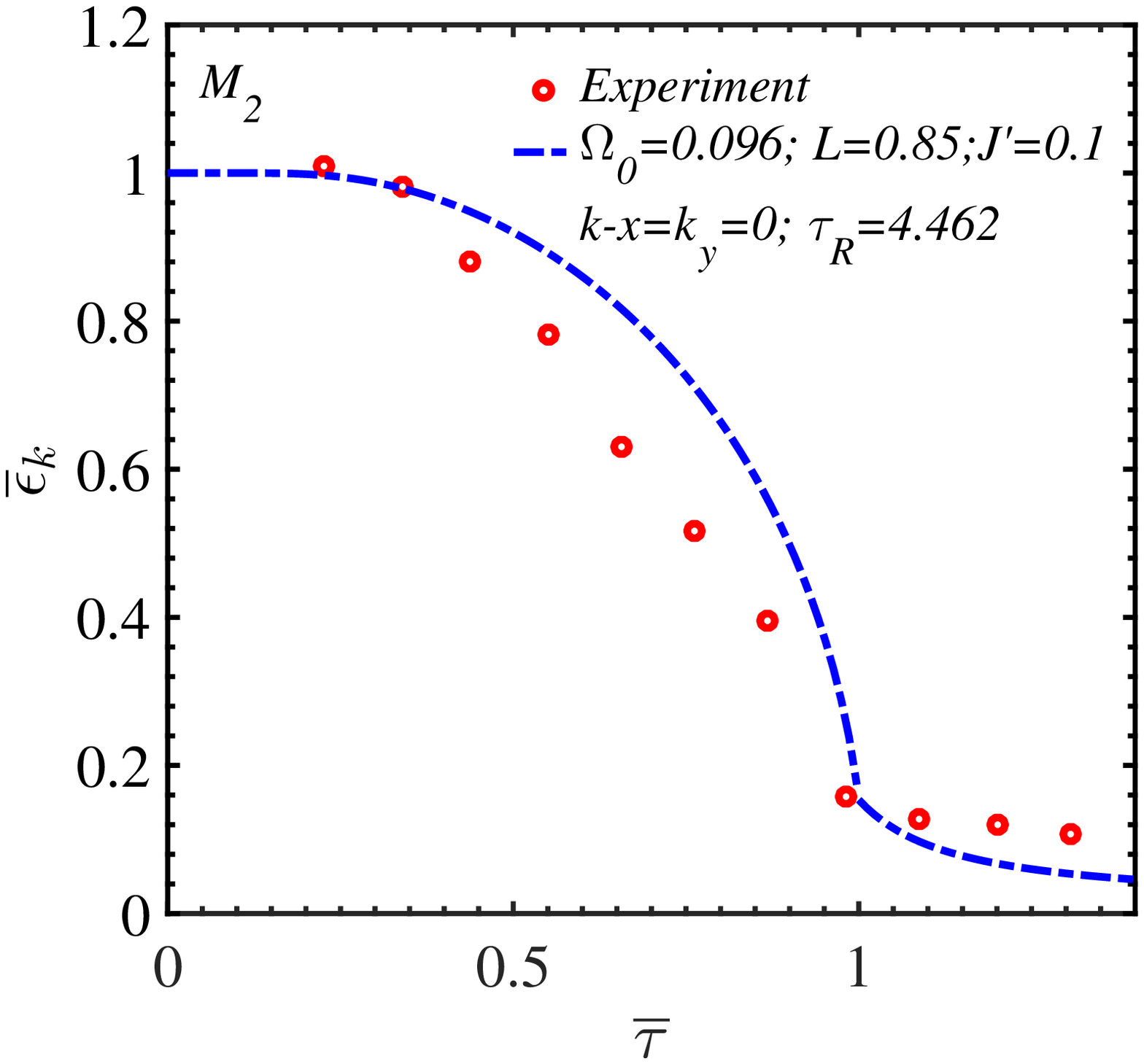} }}%
\caption{The comparison between theory and experiment \cite{He2018} for the temperature dependence of the zero-momentum magnon mode M$_1$ \protect\subref{fig:9a} and M$_2$ \protect\subref{fig:9b}. The quantities $\bar{\epsilon}_{\bf k}$, $\bar{\tau}$ are relative magnon energy and relative temperature (see the text). Fitted parameters and the SR temperature $\tau_R$ are given in terms of $J$.}
\label{Fig9}%
\end{figure}
   
    A ferromagnetic order with an out-of-plane spin orientation was experimentally observed in the CrI$_3$ monolayer. The temperature dependence of the integrated intensity of the M$_1$ and M$_2$ magnon modes performed in Ref.~\cite{He2018} was carried out on 13-layer films. To apply the single layer model for this CrI$_3$ thin film, we propose an out-of-plane magnetic moment with an initial incline angle of the film plane caused by the competition between anisotropic exchanges and TF fields. The effective anisotropy originated from the transverse strain in 2D materials plays the role of the transverse field in the model. The presence of the magnon mode intensity above SR temperature is an evidence of the transverse field.  
    
Fig.~\ref{Fig9} shows the theoretical fit with the experimental data for the temperature dependence of the zero-momentum M$_1$ and M$_2$ magnon modes in CrI$_3$ \cite{He2018}. Here, the ratio of zero-momentum magnon energy at finite temperatures to this at zero temperature and relative temperature are denoted as $\bar{\epsilon_{\bf k}}=\epsilon_{\bf k}(T)/\epsilon_{\bf k}(0)$ and $\bar{\tau}=T/T_R$, respectively. The fitted theoretical parameters presented in terms of the out-of-plane exchange constant $J$ are $J’=0.1;\, L=0.85; \,\Omega_0 = 0.096; \, \tau_R=4.462$ where the SR temperature $\tau_R$ approximates to the Curie temperature $\tau_C=4.5$ estimated by Eq.~\ref{Eq38}. Using the experimental value of the SR temperature of CrI$_3$ monolayer in Ref.~\cite{He2018}, $\tau_R=45$ K, we obtain $J=0.869$ meV, the in-plane NN exchange constant $L= 0.739$ meV, the isotropic NNN exchange constant $J’=0.087$ meV and the transverse field $\Omega_0 = 0.083$ meV. We also make a comparison of these exchange parameters obtained from the DFT calculations for the CrI$_3$ monolayer \cite{Yaz2020,Yadav2020,Wang2020} with our present work and with the 13-layer film data using NN anisotropic Heisenberg model \cite{He2018}. The data of different works are listed in Table 1.

 \begin{table}[tp]
  	\caption{The magnitude of the out-of-plane NN, in-plane NN, NNN exchanges J, L, J' exchange parameters obtained by different works, respectively.}
  	\begin{center}
  		\begin{tabular}{ | c | c |c | c|c|c|}
  			\hline
  			Parameter (meV)  &  Present work & Ref.~\cite{Wang2020} & Ref.~\cite{Yaz2020} & Ref.~\cite{Yadav2020}& Ref.~\cite{He2018} \\ \hline
  			 J & 0.869   & 1.025 & 1.53 & 0.72 & 5.46    \\
  			 L & 0.739  & - & - & -& 1.36  \\
  			J' &  0.087  & 0.549 & 0.38 & - & -  \\
  			\hline
  		\end{tabular}
  	\end{center}
  \end{table} 
  
 We can see that the NN exchange parameter $J$ of our work is in good agreement with the DFT derivations in Ref.~\cite{Yadav2020,Wang2020} and the evaluation of the Ref.~\cite{He2018} is too high. Taking $\Omega_0=g\mu_BB_{0x}$ where the Bohr magneton $\mu_B=5.788 \times 10^{-5}$ eV/T and the g-factor of the ion Cr$^{3+}$ is 1.98 \cite{Saiz2019}, we estimate the transverse field for the 13-layer CrI$_3$ film, $B_{0x}=0.73$ T. This value is in the interval of the in-plane saturated field for the CrI$_3$ monolayer ($\sim 0.17$ T) and for the bulk ($\sim 2$ T) \cite{He2018}. Using the lattice constant $a_0=6.867\times 10^{-10}$ m extracted in Ref.~\cite{Sales2015} and the formula (\ref{Eq45}), we readily obtain the magnon velocity in the 2D CrI$_3$ near Dirac point and at zero temperature $v_m=1.74$ km/s. We are looking forward to experimental data to compare with this value. 

\section{Conclusion}
The temperature and the field dependence of the ferromagnetic magnon spectra in the honeycomb single-layer spin film are calculated using the anisotropic exchange XZ-Heisenberg model with the transverse field. The field and temperature dependence of the magnetization are examined within the mean-field approximation while the magnon energy is calculated by using the Gaussian approximation. The finite-temperature phase transition due to the TF is the spin reorientation transition in the monolayer honeycomb spin-lattice with the initial out-of-plane magnetization. The collective excitation follows the different dispersion relations below or above the SR temperature. The transverse-field dependence of the magnon spectrum at zero temperature corresponding to QPT is also illustrated and the gap of the zero-momentum low-energy magnon branch disappears at the critical QPT field. The model application for the magnon modes in the 2D-CrI$_3$ ferromagnet reveals a suitable agreement.

\section*{Acknowledgment}
We would like to thank the grant Nafosted 103.01-2019.324 for support. 

\bibliographystyle{elsarticle-harv.bst}

\begin{thebibliography}{9}

	\bibitem{Balatsky2014}T.~O.~Wehling, A.~M.~Black-Schaffer and A.~V.~Balatsky, {\it Adv.~Phys}.~{\bf 63}, 1 (2014).
	
	\bibitem{Balatsky2016} J.~Fransson, A.~M.~Black-Schaffer, and A.~V.~Balatsky, {\it Phys.~Rev.~B} {\bf 94}, 075401 (2016). 
	
	\bibitem{Balatsky2018}S.~Peshoguba, S.~Banerjee, J.~C.~Lashley, J.~Park, H.~Agren, G.~Aeppli, and A.V.~Balatsky, Phys.~Rev.~X {\bf 8}, 011010 (2018).
	
	\bibitem{Boyko2018}D.~Boyko, A.~V.~Balatsky, J.~T.~Haraldsen, Phys.~Rev.~B {\bf 97}, 014433 (2018).
	
	\bibitem{Huang2017}B.~Huang, G.~Clark, E.~Navarro-Moratalla et al, Nature {\bf 546}, 270 (2017).
	
	\bibitem{Novo2019}M.~Gilbertini, M.~Koperski, A.~F.~Morpurgo and K.~S.~Novoselov, Nat.~Nanotechnol.~ {\bf 14}, 408 (2019).
	
	\bibitem{Cenker2021}J.~Cenker, B.~Huang, N.~Suri et al, Nat.~Phys.~{\bf 17}, 20 (2021).
	
	\bibitem{He2018} W.~Jin, H.~H.~Kim, Z.~Ye, S.~Li, P.~Rezaie, F.~Diaz, S.~Siddiq, E.~Wauer, B.~Yang, C.~Li, S.~Tian, K.~Sun, H.~Lei, A.~W.~Tsen, L.~Zhao, R.~He, Nat.~Commun.~{\bf 9}, 5122 (2018).
	
	\bibitem{Nguyen2018} Niem T.~Nguyen, Thao H.~Pham, Giang H.~Bach, and Cong T.~Bach, Mater.~Trans.~{\bf 59}, 1075 (2018).
	
	\bibitem{Nolting2005} S.~Schwieger, J.~Kienert, and W.~Nolting, Phys.~Rev.~B {\bf 71}, 024428 (2005).
	
	\bibitem{Scott2015} J.~F.~Scott, A.~Schilling, S.~E.~Rowley, and J.~M.~Gregg, Sci.~Technol.~Adv.~ Mater.~{\bf 16}, 036001 (2015).
	
	\bibitem{Li2020} S.~Li, Z.~Ye, X.~Luo, G.~Ye, H.~H.~Kim, B.~Yang, S.~Tian, C.~Li, H.~Lei, A.~W.~Tsen, K.~Sun, R.~He, and L.~Zhao, Phys.~Rev.~X {\bf 10}, 011075 (2020).
	
	\bibitem{Yaz2020} M.~Pizzochero, O.~V.~Yazyev, J.~Phys.~Chem.~C {\bf 124}, 7585 (2020).
	
	\bibitem{Mermin1966}N.~D.~Mermin, H.~Wagner, Phys.~Rev.~Lett.~{\bf 17}, 1133 (1966).

	\bibitem{Bach2019} Cong T.~Bach, Niem T.~Nguyen, Giang H.~Bach, Jour.~Magn.~Magn.~Mater.~{\bf 483}, 136 (2019).
	
	\bibitem{Wang2020} D.~Wang and B.~Sanyal, J.~Phys.~Chem.~C {\bf 125}, 18467 (2021).

	\bibitem{Costa2020} A.~T.~Costa, D.~L.~R.~Santos, N.~M.~R.~Peres and J.~Fernandez-Rossier, 2D Mater.~{\bf 7}, 045031 (2020).

	\bibitem{Yadav2020} M.~Pizzochero, R.~Yadav, and O.V. Yazyev, 2D Mater. {\bf 7}, 035055 (2020).
	
	\bibitem{Saiz2019} C.L. Saiz, M. A. McGuire, S.R.J. Hennadige et. al., MRS Advances {\bf 4}, 2169 (2019).
	
    \bibitem{Sales2015} M. A. McGuire, H. Dixit, V. R. Cooper, and B. C. Sales, Chem. Mater. {\bf 27}, 612 (2015).
\end{thebibliography}

\end{document}